\documentclass[aps,aip,reprint,superscriptaddress]{revtex4-1}

\usepackage{graphicx}% Include figure files
\usepackage{dcolumn}% Align table columns on decimal point
\usepackage{bm}% bold math
\usepackage{setspace}
\usepackage{color}
\usepackage{amsmath}
\usepackage{times}% Times font
\usepackage{tabularx}
\usepackage{amssymb}
\usepackage{amsfonts}
\usepackage{xcolor,colortbl}
\usepackage{booktabs}

\begin{document}

\preprint{AIP/123-QED}

\title[Sample title]{Electrical spectroscopy of forward volume spin waves in perpendicularly magnetized materials}

\author{M.~Sushruth}
\affiliation{Universit\'e Paris-Saclay, CNRS, Centre de Nanosciences et de Nanotechnologies, 91120, Palaiseau, France}
\author{M.~Grassi}
\affiliation{Universit\'e de Strasbourg, CNRS, Institut de Physique et Chimie des Mat\'eriaux de Strasbourg, UMR 7504, France}
\author{K.~Ait-Oukaci}
\affiliation{Institut Jean Lamour, CNRS, Universit\'e de Lorraine, Nancy, France}
\affiliation{SOLEIL Synchrotron, L'Orme des Merisiers, Saint Aubin - BP 48, 91192 GIF-SUR-YVETTE, France}
\author{D.~Stoeffler}
\affiliation{Universit\'e de Strasbourg, CNRS, Institut de Physique et Chimie des Mat\'eriaux de Strasbourg, UMR 7504, France}
\author{Y.~Henry}
\affiliation{Universit\'e de Strasbourg, CNRS, Institut de Physique et Chimie des Mat\'eriaux de Strasbourg, UMR 7504, France}
\author{D.~Lacour}
\affiliation{Institut Jean Lamour, CNRS, Universit\'e de Lorraine, Nancy, France}
\author{M.~Hehn}
\affiliation{Institut Jean Lamour, CNRS, Universit\'e de Lorraine, Nancy, France}
\author{U.~Bhaskar}
\affiliation{Universit\'e Paris-Saclay, CNRS, Centre de Nanosciences et de Nanotechnologies, 91120, Palaiseau, France}
\author{M.~Bailleul}
\affiliation{Universit\'e de Strasbourg, CNRS, Institut de Physique et Chimie des Mat\'eriaux de Strasbourg, UMR 7504, France}
\author{T.~Devolder}
\affiliation{Universit\'e Paris-Saclay, CNRS, Centre de Nanosciences et de Nanotechnologies, 91120, Palaiseau, France}
\author{J.-P.~Adam}
\email{jean-paul.adam@u-psud.fr}
\affiliation{Universit\'e Paris-Saclay, CNRS, Centre de Nanosciences et de Nanotechnologies, 91120, Palaiseau, France}

\begin{abstract}
We study the potential of all-electrical inductive techniques for the spectroscopy of propagating forward volume spin waves. We develop a one-dimensional model to account for the electrical signature of spin-wave reflection and transmission between inductive antennas and validate it with experiments on a perpendicularly magnetized Co/Ni multilayer. We describe the influence of the antenna geometry and antenna-to-antenna separation, as well as that of the material parameters on the lineshape of the inductive signals. For a finite damping, the broadband character of the antenna emission in the wave vector space imposes to take into account the growing decoherence of the magnetization waves upon their spatial propagation. The transmission signal can be viewed as resulting from two contributions: a first one from propagating spin-waves leading to an oscillatory phase of the broadband transmission coefficient, and another one originating from the distant induction of ferromagnetic resonance because of the long-range stray fields of realistic antennas. Depending on the relative importance of these two contributions, the decay of the transmitted signal with the propagation distance may not be exponential and the oscillatory character of the spin-wave phase upon propagation may be hidden. Our model and its experimental validation allow to define geometrical and material specifications to be met to enable the use of forward volume spin waves as efficient information carriers.
\end{abstract}

\maketitle

\section{Introduction}\label{Sec_introduction}
The eigenexcitations of magnetic materials -- the spin waves (SWs)~\cite{Chumak2015} -- are attractive for future wave-based-computing applications~\cite{Vlaminck2008,Wagner2016,Chumak2015,Vogt2014,Khitun2010,Klingler2015} because they combine small wavelengths, GHz frequencies, and tunability~\cite{Grundler2015}. Information can be stored in the amplitude and phase of the SWs~\cite{Schneider2008} and transported through a conduit made from a thin magnetic material. In in-plane magnetized thin films, the SWs are categorized depending on the orientation of their wave vector $\bm{k}$ with respect to the magnetization $\bm{M}$. The most popular configuration is the Damon-Eshbach (DE) one, where $\bm{M}\perp\bm{k}$; these waves can be efficiently generated by standard inductive or rf-spin-orbit-torque (SOT) techniques~\cite{Demidov2015,Talmelli2018}. Beside, they possess large group velocities and can therefore propagate over long distances before being attenuated. Unfortunately, the DE configuration is somewhat inadequate for transmitting spin waves in a curved conduit, because the conduit would need to be magnetized transversely by a non-uniform applied field~\cite{Haldar2016}. The slower backward volume (BV) spin waves, where $\bm{M}\parallel\bm{k}$, are far less often used \cite{bhaskar_backward_2020} because their weak coupling with inductive antennas and SOT antennas renders them difficult to excite and detect.

To enable spin-wave based transmission of information in any arbitrary direction, one can rather harness isotropic spin waves like the forward volume spin waves~\cite{Kalinikos1986} (FVSW), whose wave vector lies in the plane of a film magnetized in the out-of-plane direction. Such waves with isotropic propagation recently enabled logic operations~\cite{Klingler2015}. Unfortunately, most past studies of FVSW~\cite{Vlaminck2008} relied on materials with easy-plane anisotropy and therefore required the application of an unpractical strong perpendicular magnetic field, substantially diminishing the benefits of the FV configuration. In addition, most of the studies on FVSW~\cite{Klingler2015, Kanazawa2016, Chen2018} used ferrites, i.e. materials which are hardly portable to the silicon platform and suffer from a low saturation magnetization $M_s$, synonymous of a modest spin-wave group velocity, limiting the ability to transport information in a fast manner. A priori, materials with both a perpendicular magnetic anisotropy (PMA) and a large magnetization-thickness product would be much preferred for FVSW applications~\cite{Han2019}. This qualitative speculation needs however to be backed up by predictive models.

In this paper, we study whether transition-metal-based materials with perpendicular magnetic anisotropy (PMA) are adequate for forward volume spin-wave based information transport. We first develop a one-dimensional analytical model to account for the electrical signature of spin-wave reflection and transmission between inductive antennas. This analytical model sheds light on two particular limiting cases of interest when manipulating spin waves in inductive transceivers configuration: i) when the spin-wave attenuation length is much longer than the characteristic size of the antennas and ii) when it is much shorter.  We then discuss the exact influence of the geometry (dimensions and separation between the antennas) and of the material damping parameter on the spin-wave signals expected in the general case. We finally report on experimental investigations of Co/Ni multilayers, which validate the model. Our findings promise to be insightful for the edition of the geometrical and material specifications to be met to enable the use of FVSW as efficient information carriers.

%%%%%%%%%%%%%%%%%%%%%%%%%%%%%%%%%%%%%%%%%%%%%%%%%%%
\section{Analytical study of the spin-wave signals in inductive transceivers configurations}\label{Sec_Analytics}
In this section, we develop a simple analytical model to describe the electrical signals encountered when exciting and detecting FVSW with inductive antennas of canonical geometries. This model can be viewed as a simplified version of the formalism developed by Vlaminck and Bailleul~\cite{Vlaminck2008, Vlaminck2010}. Here, our objective is to maintain the formalism at a sufficiently didactic level so as to ease the interpretation of the lineshapes of the spin-wave inductive signals. We have systematically checked that the present model yields the same results as the more complicated model established in Refs.~\onlinecite{Vlaminck2008} and \onlinecite{Vlaminck2010}.

In the present study, we assume that the width of the spin-wave conduit is much larger that any other dimension of the system and that the inductive antennas are infinitely long in the transverse direction $y$. Consequently, we consider that all relevant magnetic fields, including the applied rf field and the demagnetizing field, are sufficiently uniform in the transverse direction so that we can restrict our analysis to the sole spin waves with a wave vector purely oriented in the longitudinal direction $x$. We shall omit the $x$ subscript and systematically write $k$ instead of $k_x$. We examine the response to a continuous-wave excitation at the working frequency $\omega_0$ and we aim to describe the lineshapes of the transmission signal between two inductive antennas, $\tilde{T}(\omega_0)$, and that of the reflection signal of each individual antenna, $\tilde{R}(\omega_0)$.

%%%%%%%%%%%%%%%%%%%%%%%%%%%%%%%%%%%%%%%%%%%%%%%%%%%
\subsection{Susceptibility for each wave vector}\label{Sec_chi_vs_k}
In spin-wave spectroscopy experiments, we work in the \textit{forced} oscillation regime and impose a harmonic stimulus at the working frequency $\omega_0$. In response to this stimulus, the magnetic system is excited within a broad range of wave vectors: It naturally responds at the wave vector $\pm k_0$ (with $k_0$ conventionally chosen positive) determined by the spin-wave dispersion relation but it is also susceptible to respond at wave vectors $k$ different from $\pm k_0$. For the sake of clarity, we shall refer to a forced oscillation response at $|k| \neq k_0$ as a \textit{magnetization wave}, in opposition to the (resonant) \textit{spin-wave} at $k_0$. The amplitude and phase of each magnetization wave $\{\omega_0,k\}$ created in response to the inductive excitation torque are governed by a complex tensorial susceptibility $\chi(\omega_0,k)$. For FVSW with weak damping and negligible exchange, the in-plane susceptibility to an in-plane harmonic field $h_x$ reads (see appendix I)
\begin{equation}
    \chi_{xx} (\omega_0,k) \approx \frac{\gamma_0^2 M_{s} H_z^\textrm{eff}}{(\omega_{k}^2 - \omega_0^2) + \imath  \omega_0 \Delta\omega_{k}\;  }.
	\label{Eq_chi_vs_k}
\end{equation}
Here, $\gamma_0$ is the gyromagnetic ratio, $H_z^\textrm{eff}$ is the equilibrium effective field, and $\omega_k$ is the spin-wave dispersion relation describing the frequency of the \textit{free} oscillations of magnetization at a given wave vector
\begin{equation}
    \omega_k^2=\gamma_0^2  {{H_{z}^\textrm{eff}} \, ({H_{z}^\textrm{eff}} + n(k) M_s )},
	\label{Eq_dispersion}
\end{equation}
with $n(k) = 1-\frac{1-e^{-\left| k\right| t}}{\left| k\right| t}$. Finally, $\Delta \omega_k$ is the mode linewidth, which can be expressed as
\begin{equation}
    \Delta \omega_k = \alpha \gamma_0 (2H+ n(k) M_s).
\end{equation}

The longitudinal susceptibility $\chi_{xx}(\omega_0,k)$ is essentially a Lorentzian distribution with a finite width determined by damping. At fixed wave vector, its full width at half maximum in the frequency space is, by definition, $\Delta \omega_k$. At fixed frequency, its pendant in the wave vector space is $\Delta k_\textrm{Gilbert}= 2/ L_\textrm{att}$, where $L_\textrm{att} = 2 |v_g(k)| / \Delta \omega_k$ is the attenuation length and $v_g \equiv \frac{\partial \omega_k}{\partial k}$ is the group velocity. For FVSW, in the small wave vector limit ($n(k) \approx \frac{\left| k\right| t}{2}$), the latter writes
\begin{equation}
    \left| v_g(k) \right| \approx \gamma_0 \frac{  {H_z^\textrm{eff}} {M_s} t}{4 \sqrt{H_z^\textrm{eff} \left(H_z^\textrm{eff} + \frac{{M_s} t |k| }{2}\right)}} \approx  \frac{\gamma_0 M_s t}{4},
    \label{Eq_vg}
\end{equation}
from which it comes~\cite{Gladii2016}
\begin{equation}
    L_\textrm{att}(k) \approx \frac{1}{\alpha}\frac{\gamma_0 M_s}{4 \omega_\textrm{FMR}} t.
    \label{Eq_Latt}
\end{equation}
Equation~\ref{Eq_vg} recalls that, at reasonably small wave vectors, the FVSW are essentially non-dispersive and that materials with large saturation magnetization are desirable to achieve large group velocities.

%%%%%%%%%%%%%%%%%%%%%%%%%%%%%%%%%%%%%%%%%%%%%%%%%%%
\subsection{Efficiency of inductive antennas in reciprocal space}\label{Sec_efficiency_function}
\begin{figure}
\includegraphics[width=8.6cm]{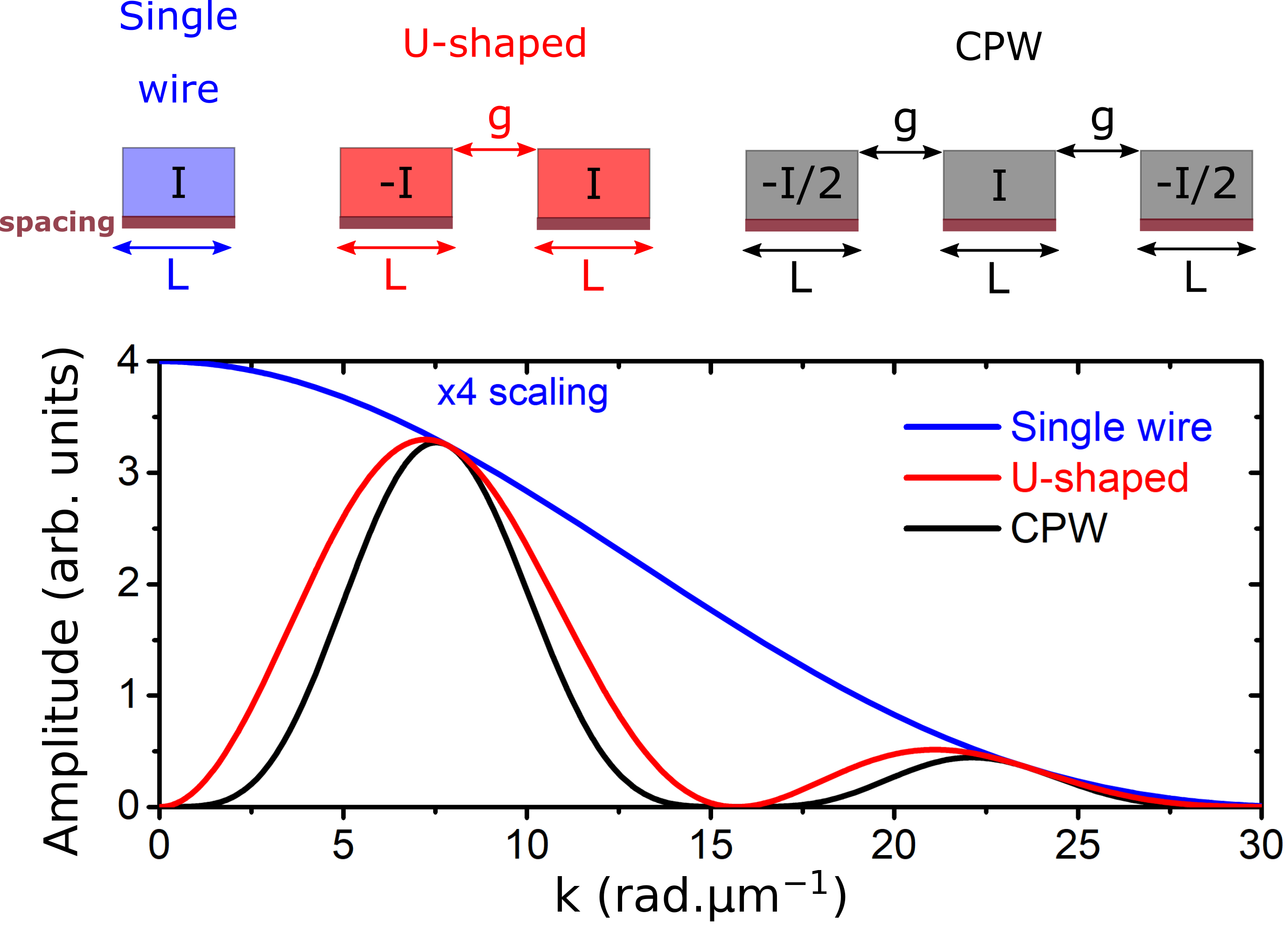}
\caption{Square modulus of the antenna efficiencies (Eqs.~\ref{Eq_AeffS}-\ref{Eq_AeffGSG}) for three different types of antennas: single wire (blue), U-shaped (red) and CPW (black), where $L = g = 200~\textrm{nm}$ and the antenna to spin wave conduit spacing is $s=0$. The single wire curve is plotted after multiplication by a factor of 4.}
\label{Fig_anteff}
\end{figure}

Since each magnetization wave $\{\omega_0,k\}$ has a finite susceptibility to the in-plane harmonic field components, it will only be excited if a stimulus with the appropriate spatial periodicity $2\pi/k$ is present; this depends solely on the antenna geometry. We consider below the three simplest antenna geometries (Fig.~\ref{Fig_anteff}), namely single wire antenna (labelled S), U-shaped antenna (labelled GS) and coplanar waveguide antenna (labelled GSG). We assume that all antennas are infinitely thin and that each conductor carries a uniform current density. Furthermore, we assume that no Eddy current is induced in any other part of the sample. Under these conditions, the $x$ component of the Oersted field for a single-wire antenna of width $L$, separated from the magnetic medium by a vertical spacing $s$, takes the form of a damped Sinc function in the wave vector space~\cite{2009a}:
\begin{equation}
	h_{x}^\textrm{S}(k) \propto \frac{\textrm{sin}(k L/2)}{k L/2} e^{-|k| s}.
	\label{Eq_AeffS}
\end{equation}
This means that, in real space, $h_{x}$ is essentially constant under the antenna and almost vanishes everywhere else, decaying in a power-law manner.

The Oersted field created by composite antennas (GS and GSG) may be deduced from Eq.~\ref{Eq_AeffS} using trivial summations. For a U-shaped antenna of gap $g$ (Fig.~\ref{Fig_anteff}), it is
\begin{equation}
	h_{x}^\textrm{GS}(k) = 2\imath ~ \textrm{sin}\!\left( \frac{k(g+L)}{2} \right) h_{x}^\textrm{S}(k),
	\label{Eq_AeffGS}
\end{equation}
and for the coplanar waveguide antenna with the same gap, it becomes
\begin{equation}
	h_{x}^\textrm{GSG} (k) = 2 ~ \textrm{sin}^{2}\!\left(\frac{k (g+L)}{2}\right) h_{x}^\textrm{S}(k).
	\label{Eq_AeffGSG}
\end{equation}

%%%%%%%%%%%%%%%%%%%%%%%%%%%%%%%%%%%%%%%%%%%%%%%%%%%
%%%%%%%%%%%%%%%%%%%%%%%%%%%%%%%%%%%%%%%%%%%%%%%%%%%

Fig.~\ref{Fig_anteff} gathers the antenna efficiency functions (Eqs.~\ref{Eq_AeffS}-\ref{Eq_AeffGSG}) for representative experimental parameters. Three points are worth noticing. First, the efficiency of all antennas vanishes at $k=2 \pi / L$. Second, single-wire antennas can excite at FMR ($k=0$) whereas composite antennas cannot. Third, the spectrum of the CPW antennas is comparable but sharper than that of U-shaped antennas. At this stage, it is useful to define a characteristic dimension $L_\textrm{ant}$ illustrating the spectral spread of the antenna. We define it as the inverse of the full width at half maximum of the main peak in the antenna efficiency function (Fig.~\ref{Fig_anteff}).

%%%%%%%%%%%%%%%%%%%%%%%%%%%%%%%%%%%%%%%%%%%%%%%%%%%
\subsection{Spin wave reflection and transmission signals}\label{Sec_spin-wave_signals}
The magnetic sample responds to the field produced by the emitter at all wave vectors present in the excitation spectrum $h_x(k)$ and, assuming that the two antennas are identical, the magnetic flux leaking from the spin-wave conduit is detected inductively by the receiver with the same transduction efficiency $h_x(k)$. As a result, the spin-wave contribution to the self-inductance of an antenna, which governs its microwave reflection coefficient~\cite{2009a}, writes
\begin{equation}
	\tilde{R}(\omega_0)= R_\textrm{norm} \int_{-\infty}^{+\infty} ~ h_{x}(k)^2 ~ \chi_{xx}(\omega_0, k)~dk,
	\label{Eq_reflexion}
\end{equation}
where $R_\textrm{norm}^{-1} = \int_{-\infty}^{+\infty} ~ h_{x}(k)^2 ~dk$ is a normalization factor.

Let $r$ be the center-to-center algebraic distance between the emitting and receiving antennas. Upon propagation, each magnetization wave undergoes a phase rotation $e^{-i k r}$. The spin-wave contribution to the mutual inductance of a pair of antennas, which governs the microwave transmission coefficient, can therefore be written as the inverse Fourier transform of $h_{x}(k)^2 \chi_{xx}(\omega_0, k)$, i.e. as
\begin{equation}
	\tilde{T}(\omega_0) = R_\textrm{norm} \int_{-\infty}^{+\infty} h_{x}(k)^2 ~ \chi_{xx}(\omega_0, k) ~ e^{-\imath kr} ~ dk.
	\label{Eq_transmission}
\end{equation}

We emphasize that the magnetic flux detected by the receiving antenna is the sum of contributions from \textit{all} excitations, not only from the resonant $k_0$ spin wave but also from all other excited magnetization waves, among which some have wave vectors with sign opposite to that of $r$. We note also that the Gilbert losses are fully accounted for in the \textit{complex} susceptibility (Eq.~\ref{Eq_chi_vs_k}). Therefore, there is no need to introduce an explicit loss term of the form $e^{-\frac{|r|}{L_{att}}}$ in Eq.~\ref{Eq_transmission}. As we shall see below, the latter appears naturally upon integration of Eq.~\ref{Eq_transmission} in cases where the inductive signal is dominated by the contribution from the resonant spin wave $\{\omega_0,k_0\}$. This justifies the common practice~\cite{Ciubotaru2016,Talmelli2018,Talmelli2019} of introducing such a term in an ad-hoc manner when the choice is made of considering the sole resonant spin wave and neglecting off-resonant magnetization waves.

%%%%%%%%%%%%%%%%%%%%%%%%%%%%%%%%%%%%%%%%%%%%%%%%%%%
\begin{table}[ht]{\hspace{1cm}}
\caption{Material parameters and SW characteristics}
\centering
\begin{tabular}{c c}
\hline\hline
Parameter & Value \\ [1ex]
\hline
Effective field~ $\mu_0 H_z^\textrm{eff}$  & 0.25 T  \\
Gyromagnetic ratio ~$\gamma_0/(2\pi)$ & 30 GHz/T \\
Magnetization  $\mu_0 M_{s}$ & 0.89 T  \\
Magnetic thickness~ $t$ & 20 nm  \\ % 20 or 16.8 nm ?
Group velocity~ $v_g(k)$ & 105 m/s \\
Damping constant~ $\alpha$ & 0.01  \;\; \;   -  \;\;\;    0.1 \\
SW linewidth ~ $\Delta \omega_k/(2\pi)$ & 230 MHz - 2.3 GHz  \\
Attenuation length~ $L_\textrm{att}(k)$ & 1.2 $\mu$m \;\; - \; 0.12 $\mu$m  \\
 [1ex]
\hline \hline
\end{tabular}
\label{Tab_matparam}
\end{table}
%%%%%%%%%%%%%%%%%%%%%%%%%%%%%%%%%%%%%%%%%%%%%%%%%%%

\subsection{Lineshape of the reflection signal versus size of the antenna}\label{Sec_reflection_vs_size}
\begin{figure}[htbp]
	\includegraphics[width=8.6cm]{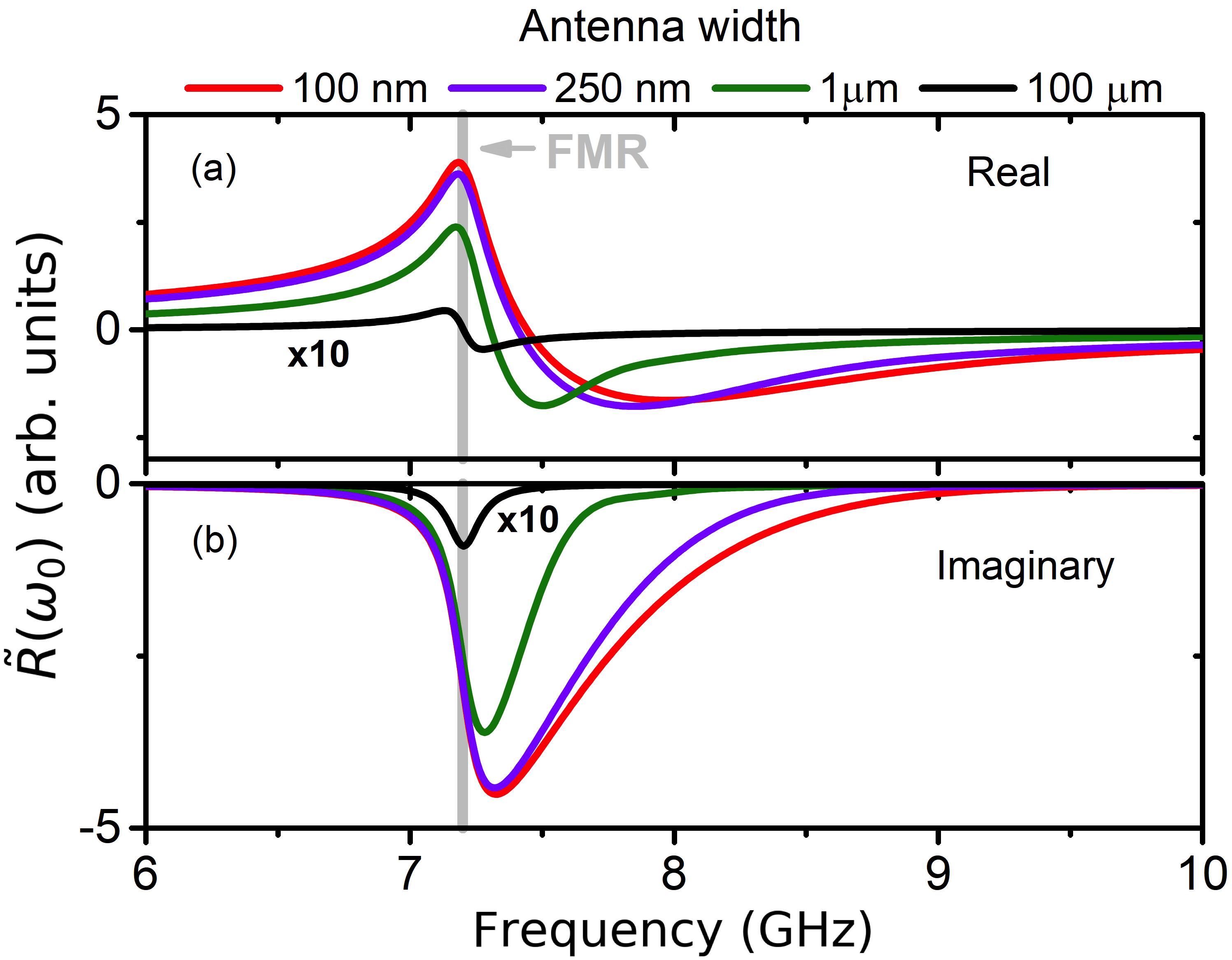}
	\caption{Modelled reflection signal $\tilde{R}(\omega)$ of single wire antennas of different widths. Real (a) and imaginary (b) parts of Eq.~\ref{Eq_reflexion}. The material parameters are that of table~\ref{Tab_matparam} with a damping chosen to be 0.01.}
	\label{Fig_noprop}
\end{figure}

Let us now examine how the antenna geometry affects the shape of the collected spin wave signals. Fig.~\ref{Fig_noprop} plots the frequency dependence of the reflection signal $\tilde{R}$ (Eq.~\ref{Eq_reflexion}) for an hypothetical single wire antenna. For large $L$ ($L \geq 100~\mu\textrm{m}$), our calculation describes a typical vector-network-analyzer(VNA)-FMR experiment using a waveguide with very wide central conductor~\cite{Devolder2013}. In this case, the antenna efficiency function is essentially restricted to a Dirac distribution around $k=0$ so that the spin-wave signal almost reduces to $\chi(\omega_0, k=0)$. Upon decreasing $L$, the antenna starts to emit in a wider band, from $k=0$ to typically $\delta k_\textrm{antenna}=\pi/L$, which yields a substantial response above FMR frequency. This wave-vector spectral spread related to the finite width of the antenna corresponds to a linewidth enhancement $\delta \omega_\textrm{antenna} = \delta k_\textrm{antenna} \times \frac{\partial \omega_k}{\partial k}$, which reads

\begin{equation}
    \delta \omega_\textrm{antenna} = \frac{\pi \gamma_0 M_s t} {4L}.
    \label{Eq_spectral_spread_1}
\end{equation}
Thus, the reflection signal $\tilde{R}(\omega_0)$ resembles the FMR susceptibility $\chi(\omega_0,k=0)$ only when the spectral spread has much less impact than the Gilbert linewidth, i.e. when $\delta \omega_\textrm{antenna} \ll \Delta \omega_\textrm{FMR}$. Equivalently, the lineshape of the reflected signal is given by the sole Gilbert damping only when $\alpha \gg \alpha_\textrm{antenna}$, where
\begin{equation}
    \alpha_\textrm{antenna} = \frac{\pi}{32} \frac{t}{L} \frac{\gamma_0 M_s}{\omega_\textrm{FMR}}.
    \label{Eq_spectral_spread_2}
\end{equation}

Note that this condition of Gilbert-dominated linewidth is not always satisfied in standard VNA-FMR characterization. For instance, if performing VNA-FMR on a 50~nm thick film at 6 GHz with antennas of width $L=50~\mu\textrm{m}$, the lineshape of the response will be almost independent from the film damping if the latter is less than $\alpha_\textrm{antenna} = 10^{-4}$. This indicates that Gilbert damping parameters smaller than a few $10^{-4}$, like those of some YIG films, cannot be properly quantified with such a setup. Conversely, for materials with larger damping, such as standard PMA transition metals (Table~\ref{Tab_matparam}), substantial distortions of the reflection signal $\tilde{R}(\omega)$ (or of the VNA-FMR signal) are not expected unless sub-micron antennas are used. We benefit from this in the experimental study of Co/Ni multilayered films, reported in section IV.

%%%%%%%%%%%%%%%%%%%%%%%%%%%%%%%%%%%%%%%%%%%%%%%%%%%
%%%%%%%%%%%%%%%%%%%%%%%%%%%%%%%%%%%%%%%%%%%%%%%%%%%

\subsection{Lineshape of the transmission signal}\label{Sec_lineshape_of_transmission}
According to Eq.~\ref{Eq_transmission}, the transmission signal is built from the interferences of various magnetization waves with wave vectors spread over an interval of width $\Delta k$, defined by the overlap of the susceptibility and antenna efficiency functions, $\chi_{xx}(\omega_0,k)$ and $h_x(k)$, respectively. Upon propagation of these waves over a distance $r$, the spread in their phases increases by an amount $\Delta k\,r$. Consequently, the interferences become more and more destructive as the propagation distance increases and the amplitude of the transmission signal decreases to eventually approach zero when the phase spread becomes much greater than $2\pi$. Mathematically, this phase decoherence occurs because the propagation operator $e^{-\imath kr}$ oscillates faster and faster and the integrand $ h_{x}(k)^2 \chi(\omega_0,k) e^{-\imath kr}$ in Eq.~\ref{Eq_transmission} averages out to a smaller and smaller value. 

A priori, the transmission signal $\tilde{T}$ involves three different lengths: the propagation distance $r$, the characteristic size of the antenna $L_\textrm{ant}$, and the spin-wave attenuation length $L_\textrm{att}$. If the conditions $r \gg L_\textrm{att}$ and $r \gg L_\textrm{ant}$ are both fulfilled, complete decoherence of the excited magnetization waves is achieved and the transmission signal completely vanishes. Besides the very peculiar case of zero damping, which yields a trivial oscillatory transmission extending to infinite distances (see Appendix II), one can identify two limiting cases in which the decoherence-induced decay of magnetization waves, hence the transmission signal, take very distinct forms: $L_\textrm{att} \gg L_\textrm{ant}$ and $L_\textrm{att} \ll L_\textrm{ant}$.

%%%%%%%%%%%%%%%%%%%%%%%%%%%%%%%%%%%%%%%%%%%%%%%%%%%
\subsubsection{Transmission signal in the limit of long attenuation length}\label{Sec_transmission_at_long_Latt}
Let us first consider the situation where the damping is small enough so that $L_\textrm{att} \gg L_\textrm{ant}$ and the spin waves can propagate far away from the emitting antenna. This corresponds to $\alpha < \alpha_\textrm{antenna}$ if single wire antennas are used. In this case, the integrand in Eq.~\ref{Eq_transmission} is dominated by the values of $k$ very close to the poles of the susceptibility, $\pm k_0$. Linearizing the spin-wave dispersion and assuming a slow variation of $v_g$ and $h_x$ with $k$ around these poles, the transmission signal may be approximated as
\begin{equation}
    \tilde{T}(\omega_0) \approx  R_\textrm{norm}~\frac{\gamma^2_0 M_s H_z^\textrm{eff}}{2 \omega_0\,v_{g}\!(k_0)} h_{x}(k_0)^2 \int_{-\infty}^{+\infty}  \frac{e^{-\imath kr}}{|k|-k_0+\frac{\imath}{L_\textrm{att}}}\,dk.
    \label{Eq_Tapprox}
\end{equation}
Applying the residue theorem and neglecting a fast decaying near-field contribution, the integral in Eq.~\ref{Eq_Tapprox} can be shown to amount to $2 \imath \pi e^{-\imath k_0 |r|} e^{-\frac{|r|}{L_\textrm{att}}}$ (See Appendix B in Ref.~\onlinecite{Kostylev2007}). It follows that, in the "resonant-spin-wave-dominated" regime, the transmission signal writes
\begin{equation}
    \tilde{T}(\omega_0, L_\textrm{att} \gg L_\textrm{ant}) \propto  ~\imath ~e^{-\imath k_0 |r|} ~e^{-\frac{|r|}{L_\textrm{att}}} ~h_{x}(k_0)^2.
    \label{Eq_Texperimentalist1}
\end{equation}
This expression recalls the empirical one used to model various experiments conducted in the Damon-Eshbach configuration~\cite{Ciubotaru2016,Talmelli2018,Talmelli2019,Collet2017,Pirro2014}. The factors $~e^{-\frac{|r|}{L_\textrm{att}}}$ and $e^{-\imath k_0 |r|}$, which appear naturally upon Fourier transforming the Lorentzian susceptibility function (Eq.~\ref{Eq_Tapprox}), illustrate, respectively, the exponential attenuation~\footnote{In a corpuscular approach, $L_\textrm{att}$ describes how an hypothetical 'pure' population of identical particles (magnons) sharing the same unique wave vector and the same unique frequency would progressively disappear upon propagation by loosing energy to the \textit{non-magnetic} degrees of freedom. However, this way of apprehending the facts is quite hypothetical since getting a 'pure' population of identical magnons would require a perfectly periodic antenna extending to both infinities, a situation where the concept of spatial attenuation cannot be defined.} and the gradual phase rotation of the spin wave upon propagation. The factor $\imath$ arises from the fact that the response at the poles is solely related to the imaginary part of the susceptibility since the real part vanishes there. Finally the antenna efficiency $h_{x}(k_0)$ matters, as expected. As demonstrated in Appendix II, Eq.~\ref{Eq_Texperimentalist1} naturally holds for $\alpha=0$.

%%%%%%%%%%%%%%%%%%%%%%%%%%%%%%%%%%%%%%%%%%%%%%%%%%%
\subsubsection{Transmission signal in the limit of short attenuation length}\label{Sec_transmission_at_short_Latt}
Let us now consider the opposite situation where damping is large and the antenna is much wider than the attenuation length ($L_\textrm{att} \ll L_\textrm{ant}$). If single wire antenna are used, this corresponds to $\alpha \gg \alpha_\textrm{antenna}$, a situation where the reflection signal looks like the FMR susceptibility (see Sec.~\ref{Sec_reflection_vs_size}). In this limit, the susceptibility $\chi_{xx}(\omega_0,k)$ is almost constant in the region where the antenna efficiency function $h_{x}(k)^2$ is contributing most to the integrand of Eq.~\ref{Eq_transmission}. Then, one can approximate the transmission signal by
\begin{equation}
    \tilde{T}(\omega_0) \approx R_\textrm{norm}\,\chi_{xx}(\omega_0,k\!=\!0) \int_{-\infty}^{+\infty} e^{-\imath kr} h_{x}(k)^2 dk.
\end{equation}
Applying Parseval-Plancherel identity, the integral in the above equation can be rewritten as $\int_{-\infty}^{+\infty} h_{x}(x)\,h_{x}(x+r)\,dx$, which measures how the inductive fields generated by two antennas separated by $r$ overlap in the volume of the magnetic medium. It thus expresses a form of inductive coupling, which does not involve the propagation of spin waves. As already noted in Ref.~\onlinecite{Birt2012}, near FMR, part of the magnetization waves existing far from the emitter can indeed be viewed as being directly excited by the long range rf field produced by the antenna. We will refer to this signal as the \textit{distant induction} of FMR.
The transmission signal in the wide antenna limit finally reads
\begin{equation}
    \tilde{T}(\omega_0, L_\textrm{att} \ll L_\textrm{ant}) \propto \chi_{xx}(\omega_0,k\!=\!0) \int_{-\infty}^{+\infty} h_{x}(x) ~h_{x}(x+r)~dx.
    \label{Eq_Texperimentalist2}
\end{equation}
This expression indicates that whenever the spin waves do not travel much, either because $L_\textrm{att}$ is too short or because the antennas are comparatively too wide, the transmission signal is mainly due to the distant induction of FMR. Therefore, in the $L_\textrm{att} \ll L_\textrm{ant}$ limit, the reflection and transmission spectra shall look alike and resemble both the FMR susceptibility.

%%%%%%%%%%%%%%%%%%%%%%%%%%%%%%%%%%%%%%%%%%%%

%%%%%%%%%%%%%%%%%%%%%%%%%%%%%%%%%%%%%%%%%%%%
\section{Numerical study of the spin-wave transmission signals}
In the general case, the three different characteristic lengths controlling the transmission spectra, i.e. $r$,  $L_\textrm{ant}$, and $L_\textrm{att}$, must all be taken into account and considered on an equal footing. Eq.~\ref{Eq_transmission} must then be integrated numerically. Below, we present results of such numerical integrations for material parameters typical of transition metal PMA films (Table~\ref{Tab_matparam}).

%%%%%%%%%%%%%%%%%%%%%%%%%%%%%%%%%%%%%%%%%%%%
\begin{figure}[htbp]
\includegraphics[width=8.6cm]{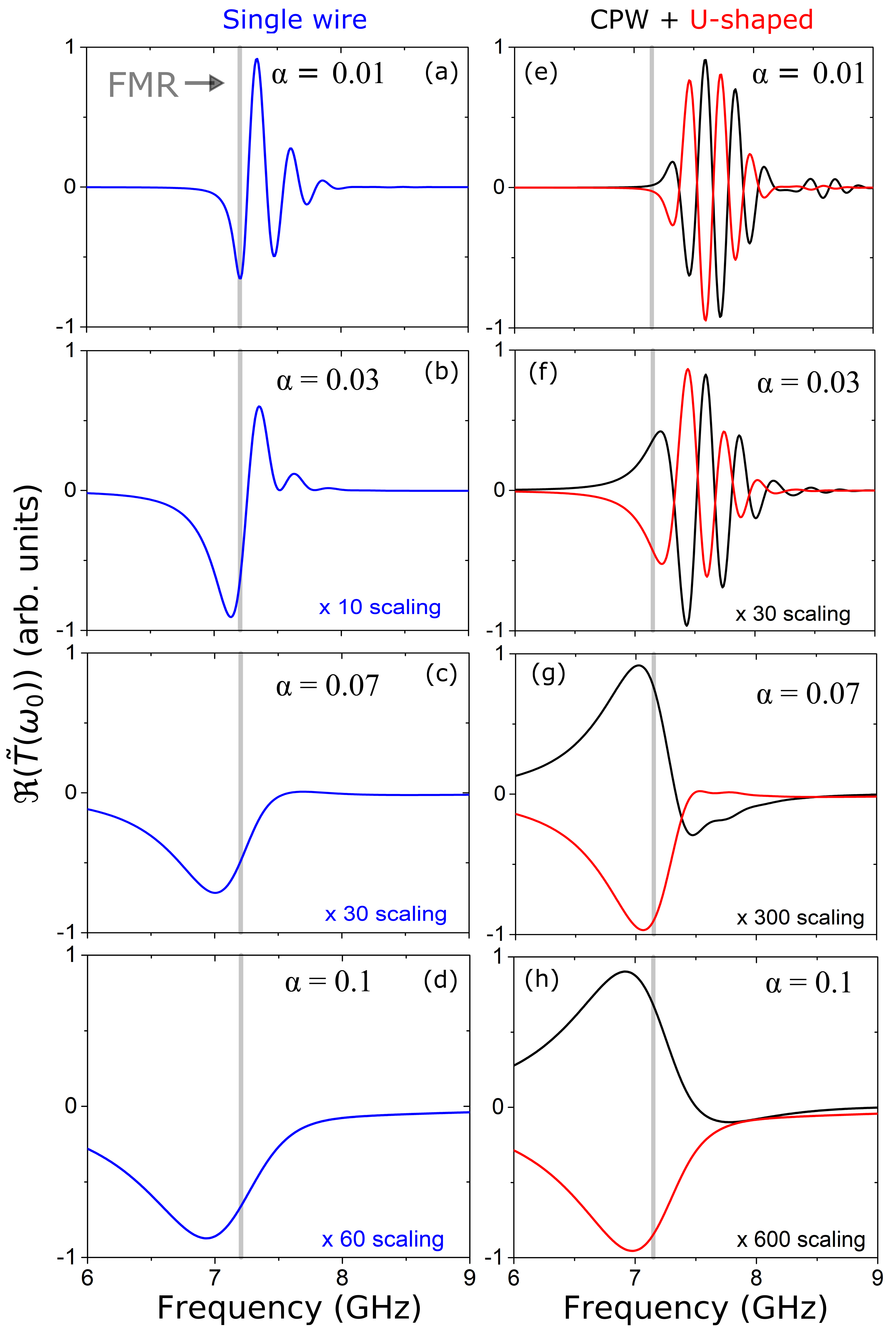}
\caption{Damping dependence of the real part of the expected transmitted signal at the receiving antenna (Eq.~\ref{Eq_transmission}) for single wire (a-d) and composite antennas (e-h). Parameters are $L=200$ nm, $g=250$ nm$, s=30$ nm and $r=2~\mu\textrm m$.}
\label{Fig_Avsalpha}
\end{figure}
%%%%%%%%%%%%%%%%%%%%%%%%%%%%%%%%%%%%%%%%%%%%

\subsection{Evolution of the transmission spectra with damping}
Let us first study how the transmission signal changes with the magnetic damping parameter, for a given propagation distance of $2~\mu$m. Figure~\ref{Fig_Avsalpha} shows the numerically determined transmission spectra $\tilde{T}(\omega_0)$ for four values of $\alpha$ and the three types of antennas introduced previously. At low damping [Fig.~\ref{Fig_Avsalpha}(a,e)], a broadband transmission is observed, in which the phase of $\tilde{T}$ rotates within an envelope defined by the antenna efficiency function (Fig.~\ref{Fig_anteff}), consisting of one (single wire antenna) or two lobes (composite antennas). These transmission spectra are dominated by the contribution of propagating spin-waves and they resemble the expectations from Eq.~\ref{Eq_Texperimentalist1}. As the damping increases [Fig.~\ref{Fig_Avsalpha}(b,c,f,g)], the transmission amplitude decreases rapidly (see the scaling factors) and its frequency span shrinks and progressively concentrate into the vicinity of FMR. Concomitantly, the oscillatory nature of the transmission signal above FMR is suppressed. The oscillations are lost at smaller damping values for single wire antennas than for composite antennas, which will be explained later in the paper. At large damping [Fig.~\ref{Fig_Avsalpha}(d,h)], the transmission spectra eventually become fully dominated by the contribution from the distant induction of the FMR and they resemble the expectations from Eq.~\ref{Eq_Texperimentalist2}.

\subsection{Evolution of the transmission spectra with distance}\label{Sec_transmission_vs_r}
Let us now examine how the transmission signal depends on the propagation distance, for a finite damping of 0.01. Figure~\ref{Fig_prop1} illustrates the case of single wire antennas, as well as CPW antennas separated by 2 to $14~\mu$m.

\subsubsection{Transmission signal below FMR}
Below FMR frequency, the transmission signal does not contain oscillations related to phase rotation upon propagation. At these frequencies, indeed, there is no spin wave mode available to transport energy. The finite transmission is exclusively due to the lateral spread of non-resonant forced oscillations (magnetization waves) and the signal originates essentially from low $k$ values, where the susceptibility is maximum. A direct consequence of this is that the transmission below FMR is extremely weak for composite antennas the efficiency of which is nil at $k=0$ (Fig.~\ref{Fig_anteff}). 

\subsubsection{Transmission signal above FMR} 
As the frequency increases above FMR, propagating spin-wave modes become accessible. Their wave vector increases with increasing frequency, following the dispersion relation (Eq.~\ref{Eq_dispersion}), which results in a frequency-growing accumulated phase rotation $-k_0 r$ after propagation over a distance $r$. The phase rotation translates into a cosine-like (resp. sine-like) variation of the real (resp. imaginary) part of $\tilde{T}$ with frequency. The frequency spacing between two successive maxima, corresponding to a phase increase of $2\pi$, may be identified with the ratio $\frac{v_g}{r}$~\cite{Gladii2016}. Here, again, the amplitude of the oscillations follows an envelope, which, at low damping, is set by the antenna efficiency function $h_x[k_0(\omega_0)]$.

\subsubsection{Crossover between resonant-spin-wave-dominated transmission and distant FMR induction}
\begin{figure}[]
    \includegraphics[width=8.6cm]{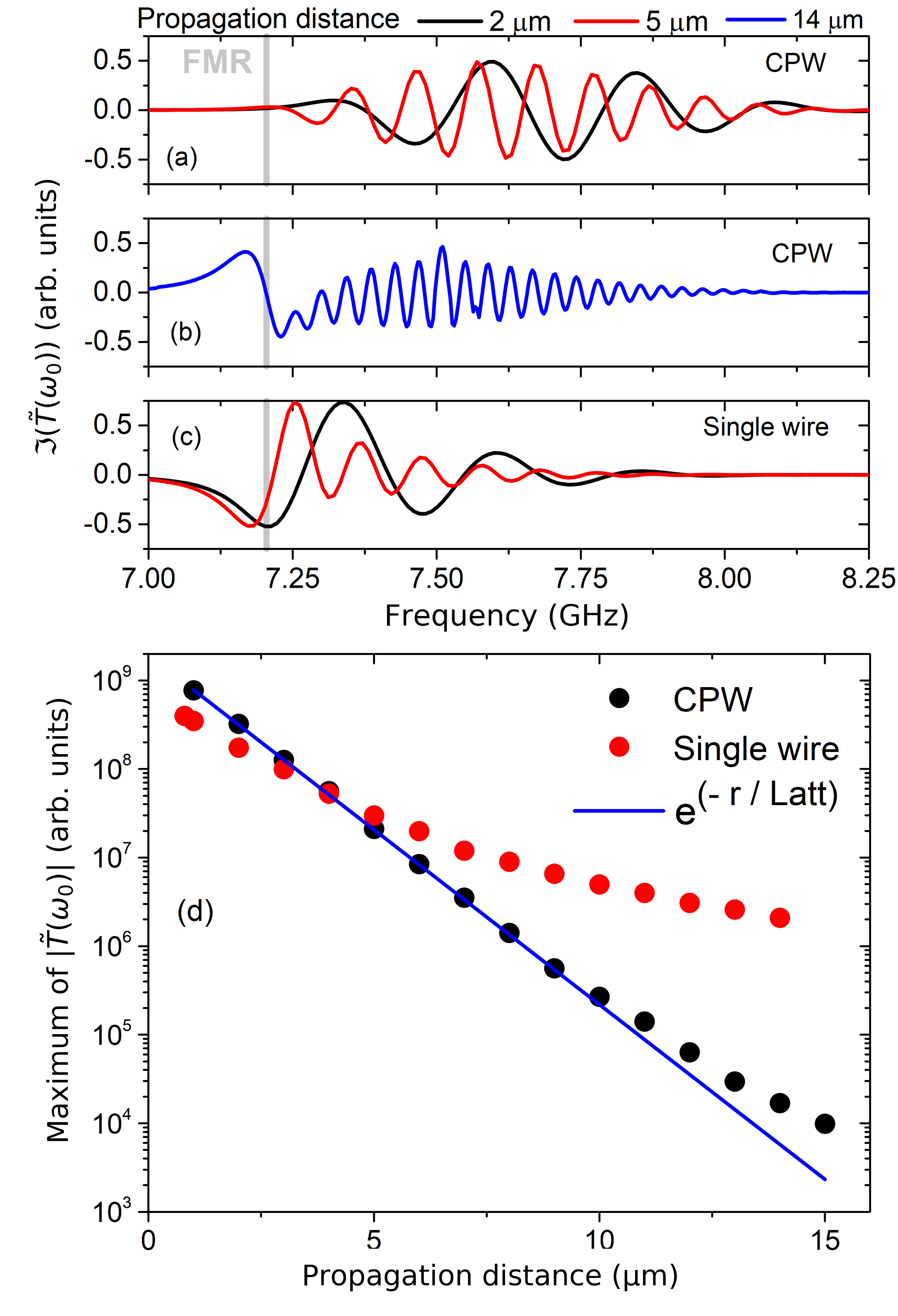}
    \caption{Influence of the propagation distance on the imaginary part of the transmission signal (Eq.~\ref{Eq_transmission}) with frequency for (a, b) CPW antennas ($L = 200$ nm, $g = 250$ nm and $s = 80$ nm) and (c) single wire antennas ($L = 500$ nm, $s = 80$ nm). The damping parameter is 0.01. Propagation distances are indicated in the legend. (d) Maximum of the modulus of the transmission coefficient versus antenna-to-antenna distance for the two antenna geometries. The blue line corresponds to an exponential decay with an attenuation length of 1.1 $\mu\textrm{m}$, as calculated from Eq.\ref{Eq_Latt}.}
    \label{Fig_prop1}
\end{figure}

Fig.~\ref{Fig_prop1}(d) shows how the maximum amplitude of the transmission signal decays with increasing distance $r$. For CPW antennas, the decay is initially exponential. Beyond some distance [$\approx 10 ~\mu\text m$ with our material parameters], however, it deviates from the commonly anticipated $e^{-\frac{|r|}{L_\textrm{att}}}$ law~\cite{Demidov2016,Collet2017,Pirro2014} and becomes significantly slower. For single-wire antennas, the decay is never truly exponential-like, even at the smallest distances. All of these deviations from a pure exponential behaviour occur because, as evoked before, the transmitted signal is generally made of two contributions, that of propagating spin waves and that of the distant induction of FMR. Since the rf field produced by the antennas decreases in a \textit{power-law} manner with distance, while propagating spin waves decay \textit{exponentially}, there will necessarily be a distance beyond which the contribution of the distant FMR induction (approximated by Eq.~\ref{Eq_Texperimentalist2}) will dominate over that of propagating spin waves (approximated by Eq.~\ref{Eq_Texperimentalist1}) and the maximum amplitude of the transmission signal will progressively tend to follow a power-law asymptote. This distance depends on the exponent of the power-law decay of the spreading rf field $h_x(r)$, hence on the type of antenna. This explains why the transition occurs at much longer propagation distance for the CPW antennas ($h_{x}^\textrm{GSG}(r) \propto r^{-4}$) than for single wire antenna ($h_{x}^\textrm{S}(r) \propto r^{-2}$) and why a much clearer crossover is observed for the former, see Fig.~\ref{Fig_prop1}(d).

%%%%%%%%%%%%%%%%%%%%%%%%%%%%%%%%%%%%%%%%%%%%
%%%%%%%%%%%%%%%%%%%%%%%%%%%%%%%%%%%%%%%%%%%%

\section{Experimental spin wave signals} 
Let us now compare the results of our model with those of travelling SW spectroscopy experiments carried out in the FVSW geometry on an archetypal metallic PMA stack, namely a Co/Ni multilayer.

\subsection{Sample} 
\begin{figure}
\includegraphics[width=8.6cm]{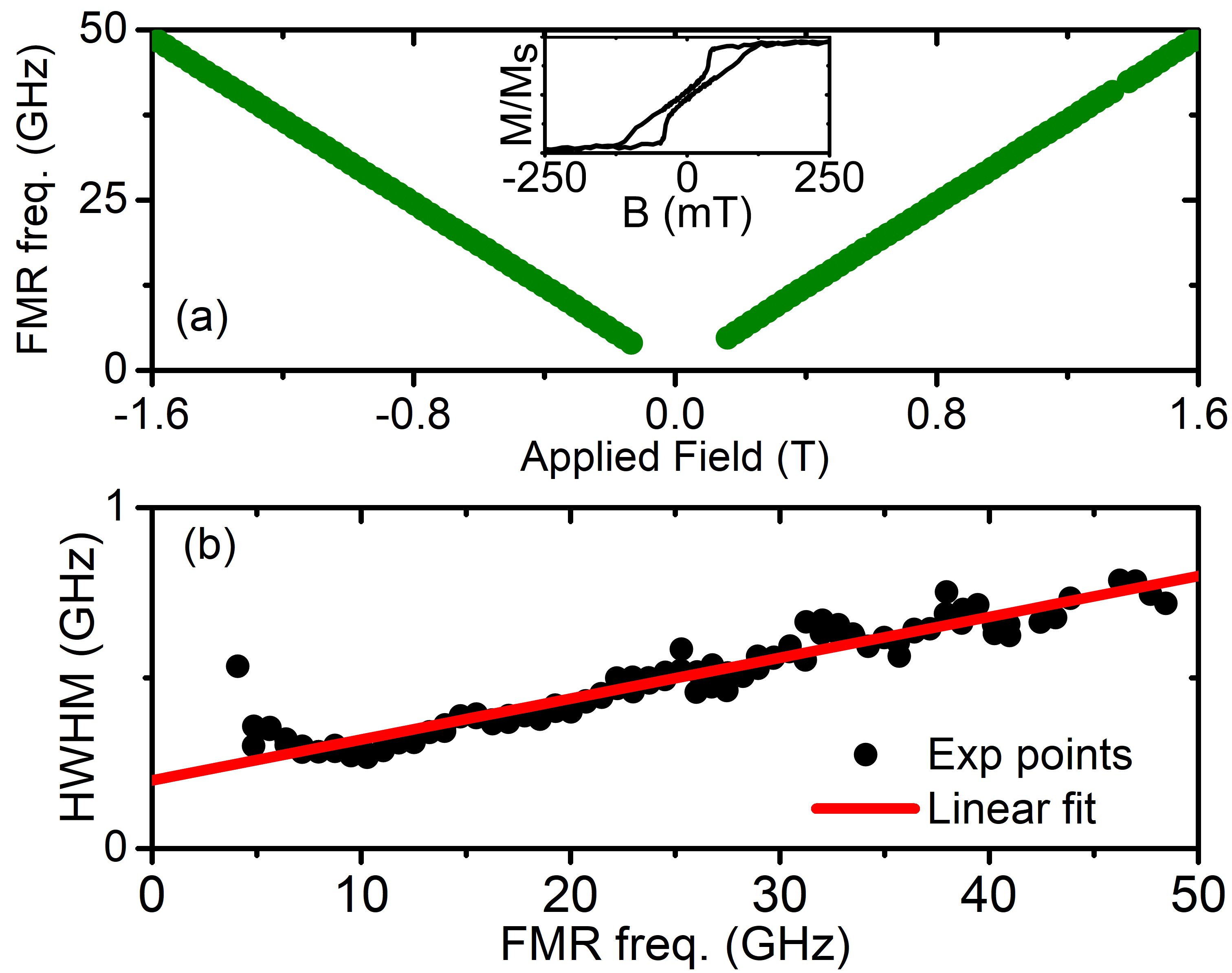}
\caption{Magnetic properties of the [Co(2\r{A})/Ni(6\r{A})$_{\times 21}$] selected multilayer. (a) Out-of-plane field dependence of FMR frequency. The inset shows the $M_z$ vs $H_z$ loop. (b) Variation of the half width at half maximum (HWHM) of the FMR line with frequency. The slope may be identified with Gilbert damping parameter. The linear fit yields $\alpha = 0.013\pm 0.002$.}
\label{Fig_exp_1}
\end{figure}

The studied device was fabricated from a multilayer of composition Pt(5)/Cu(2.5)/\{Ni(0.6)/Co(0.2)$\}_{\times 21}$/Ni(0.6)/ Cu(2.5)/Ta(2.5), where the numbers are the thicknesses of the layers in nanometers. Several such stacks were deposited on natively oxidized intrinsic high resistance Si(100) substrates by DC magnetron sputtering. Damping parameters and effective anisotropy fields were determined by conventional Vector Network Analyser-FMR in perpendicularly applied magnetic fields (Fig.~\ref{Fig_exp_1}). The obtained values span from 0.013 to 0.02, and 3 mT to 280 mT, respectively. Then the multilayer with lowest damping (0.013) and anisotropy field (3~mT) was selected for microfabrication: it was patterned into a stripe-shaped spin wave conduit by a combination of optical lithography and ion-beam etching.

%%%%%%%%%%%%%%%%%%%%%%%%%%%%%%%%%%%%%%%%%%%%
%%%%%%%%%%%%%%%%%%%%%%%%%%%%%%%%%%%%%%%%%%%%

The width of the patterned spin wave conduit (20 $\mu$m) was chosen much greater than any other dimension in the experiment so as to ensure quasi-translational invariance of the magnetic equilibrium configuration in the transverse direction and thereby promote the generation of spin waves with wave vectors directed primarily along the conduit length. The spin-wave conduit was covered with 30~nm of SiO$_{2}$ for electric insulation. Finally, the above lying 90 nm thick coplanar waveguide antennas, made of aluminum, were patterned using electron-beam lithography. The CPWs consist of conductors of widths $L=200\,\textrm{nm}$ and edge-to-edge gaps $g=250$~nm [Fig.~\ref{Fig_exp_2}(a)]. The antenna center to antenna center distance was set to $r=2\,\mu\textrm{m}$.

\subsection{Spin wave transmission signal}
The spin wave transmission signal was measured using a 2-port Vector Network Analyser (VNA). A standard phase and amplitude transmission calibration procedure of the VNA was applied in the reference planes of the device pads. Around the FMR frequency, the cross-talk between the input and output antennas was typically -70 dB. This field independent parasitic signal was larger than the spin-wave signal and we removed its contribution by subtracting an appropriate reference measured in zero field, while the sample was in a magnetic multidomain configuration and no clear spin-wave related signal could be detected.

%%%%%%%%%%%%%%%%%%%%%%%%%%%%%%%%%%%%%%%%%%
\begin{figure}
\includegraphics[width=8.6 cm]{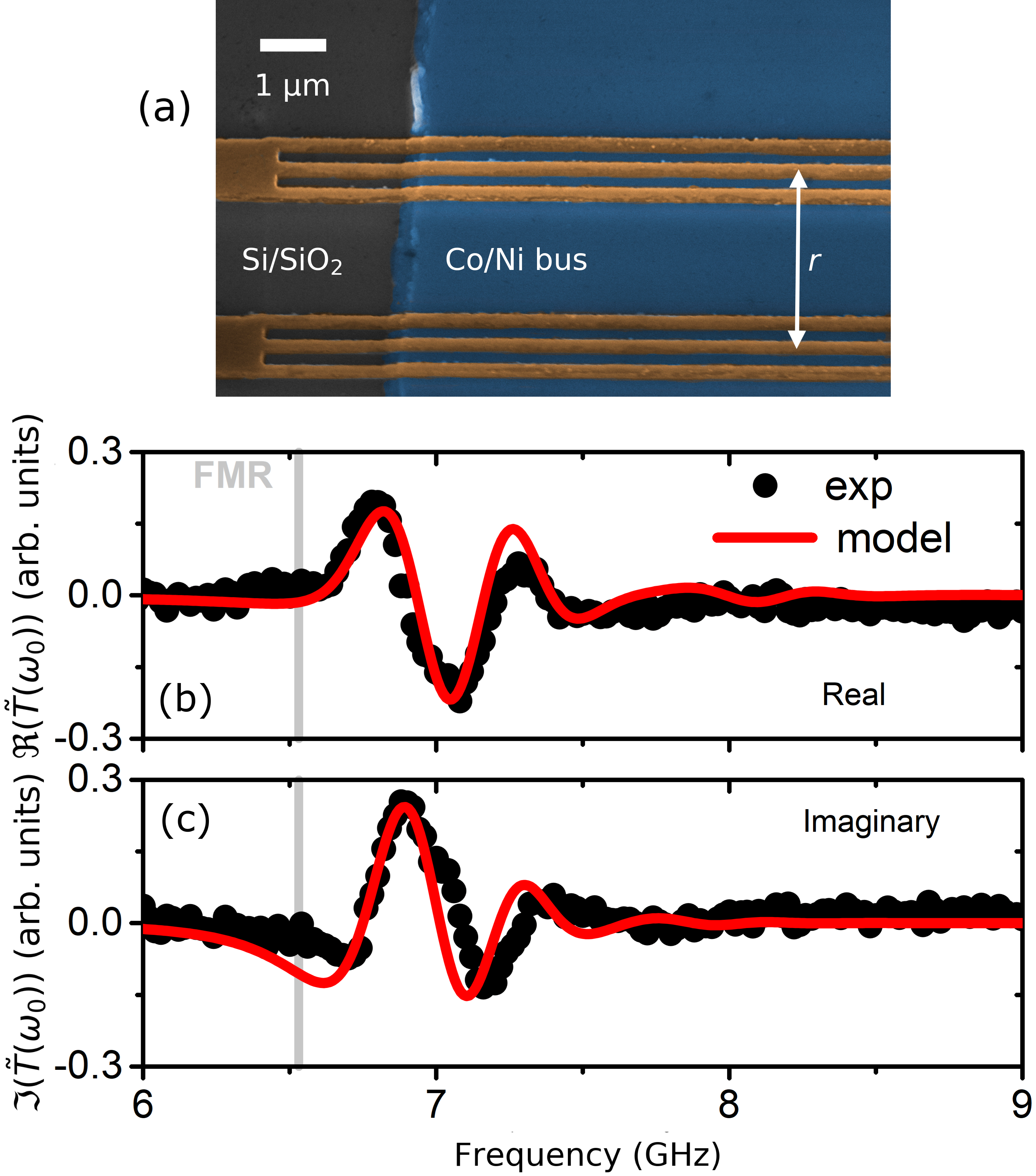}
\caption{(a) Colorised scanning electron micrograph of the device with CPW antennas on a Co/Ni multilayer. The scale bar is $1~\mu\textrm{m}$. The nominal width of the conductor $L$ is 200 nm and the gaps between the conductors are 250 nm. (b) Real and (c) imaginary parts of the experimental transmission parameter and fits thereof with a damping of 0.03, an antenna size of $L=180$ nm, a gap of $g=200$ nm, spacing of $s=80$ nm and a propagation length of 1.8~$\mu\text m$.}
\label{Fig_exp_2}
\end{figure}
%%%%%%%%%%%%%%%%%%%%%%%%%%%%%%%%%%%%%%%%%%

A representative spin-wave transmission spectrum is plotted in Fig.~\ref{Fig_exp_2}(b,c). Its main qualitative features, namely its oscillatory nature, the quadrature relation between its real and imaginary parts, the presence of a faint second lobe envelope at high frequency, and the quasi-absence of signal below FMR, are all in line with expectations. From a qualitative point of view, the agreement between experimental and theoretical spectra can be regarded as satisfactory, hence confirming the validity of the model. However, achieving a quantitative agreement between experiment and modeling required one to adapt the damping parameter $\alpha$ and/or the vertical spacing $s$ between the antennas and the magnetic material. These two parameters have a similar influence on the lineshape of the transmission parameters (see appendix III) so that they cannot be evaluated independently from each other in a reliable manner from the fitting procedure. To adjust $\alpha$, we have chosen to set $s$ as the distance separating the middle of the magnetic film and the middle of the antenna. Under this assumption, the device damping needed to be increased to 0.03 to get a satisfactory fit. This is to be compared to the value of $\alpha=0.013$ extracted from the FMR data taken on the multilayer from which the device was fabricated [Fig.~\ref{Fig_exp_1}(b)]. We believe that the observed higher damping at device level could indicate that the fabrication process may have degraded the material damping parameter, which likely correlates to the formerly observed strong variability of the damping measured on multilayers from different batches. Alternatively, it is also possible that the inhomogeneous broadening of ferromagnetic resonance, which appears in VNA-FMR as an extrapolated non-zero linewidth at zero frequency, also contributes to the effective spin-wave damping. This could be the case if the magnetic inhomogeneities at play occurred on a length scale much smaller than the dimensions of our device.

%%%%%%%%%%%%%%%%%%%%%%%%%%%%%%%%%%%%%%%%%%
\section{Conclusion}
In summary, we have investigated to what extent inductive techniques can be used for the spectroscopy of forward volume spin waves in metallic films with perpendicular magnetic anisotropy. We have developed and validated a simple model that describes the emission (Eq.~\ref{Eq_reflexion}) and the distant detection (Eq.~\ref{Eq_transmission}) of plane waves in a one-dimensional magnetic conduit with out-of-plane magnetization. The excitation and collection efficiencies in wave vector space depend on the antenna geometry: a broadband emission starting from ferromagnetic resonance ($k=0$) can be obtained with a single wire antenna, while U-shaped and coplanar waveguide antennas emit in a narrower band that excludes the vicinity of FMR. As a result of this wave vector spectral spread, the reflection signal of an antenna can have a complicated frequency-lineshape. The lineshape mimics the classical Lorentzian-like macrospin susceptibility only when the damping contribution to the linewidth dominates the contribution related to the antenna spectral spread (Eq.~\ref{Eq_spectral_spread_2}). This criterion is generally not satisfied for YIG-based conduits with micron-sized antennas but it can be met easily for PMA transition metal magnets that possess a substantially larger Gilbert damping.

The antenna-to-antenna transmission signal (Eq.~\ref{Eq_transmission}) can be understood as a combination of the wave-vector-resolved magnetic film susceptibility, the antenna spectral efficiency, as well as the magnetization-wave phase rotation upon propagation. For sufficiently low damping, the combination of the finite group velocity and the spin-wave phase rotation results in an oscillatory character of the transmission parameter versus the frequency. The damping decreases the susceptibility and mixes the contributions of the different wave vectors, which leads to the progressive decoherence of the propagating spin-wave packets and has dramatic qualitative and quantitative impacts on the transmission signal. In the limit of a moderate damping / moderate distance, this reduces to an exponential decay of the transmission signal, with a decay rate identified as the inverse of the spin-wave attenuation length. In other situations, however, the decay may be slower than exponential. Indeed, part of the transmission can be viewed as a distant induction of FMR, arising from the long-range stray fields produced by the antennas. Our findings offer physical insight that can help to engineer the geometrical and material specifications to be met to enable the use of forward volume spin waves as efficient information carriers.

\section{Acknowledgements}
The authors acknowledge financial support from the French National Research Agency (ANR) under Contract No. ANR-16-CE24-0027 (SWANGATE). U.B acknowledges in addition the FETOPEN-01-2016-2017 [FET-Open research and innovation actions (CHIRON project: Grant agreement ID: 801055)]. The authors acknowledge
the STnano cleanroom facility for technical support and Mikhail Kostylev for constructive discussions.

%%%%%%%%%%%%%%%%%%%%%%%%%%%%%%%%%%%%%%%%%%%%%%%%%%%
\section*{Appendix I: susceptibility versus wavevector and frequency}
In this first appendix, our aim is to determine the transverse susceptibility versus angular frequency $\omega \in \mathbb{R}$ and wave vector $k\in \mathbb{R}$ of a PMA film in response to a transverse rf magnetic field. The calculated susceptibility is used to deduce the shape of the line expected in the forward volume wave spectroscopy models and experiments presented in this paper. 

We calculate the dynamic response of the magnetization $\bm{M}$ in the presence of a magnetic field $\bm{H}$ starting from the Landau-Lifshitz-Gilbert (LLG) equation:
\begin{equation}
    \frac{\partial \bm{M}}{\partial t} = -\gamma_0 \left( \bm{M} \times \bm{H} \right) + \frac{\alpha}{M_\textrm{S}} \left(\bm{M} \times \frac{\partial \bm{M}}{\partial t}\right).
    \label{Eq_LLG_1}
\end{equation}
Here, $\gamma_0= \gamma \, \mu_0$, where $\gamma$ is the gyromagnetic factor, $\mu_0$ is the permeability of vacuum, $M_\textrm{s}$ is the saturation magnetization, and $\alpha$ is the dimensionless damping parameter. 

In order to linearize the above equation around the equilibrium $\bm{M}//z$, we may decompose the magnetization vector as the sum of static and dynamic components
\begin{equation}
    \bm{M}(t)=M_s (\bm{e_{z}} +\bm{\delta m}(t)),
    \label{mag}
\end{equation}
where the dynamic magnetization $\bm{\delta m}(t)$ is assumed to have the form of a plane wave $\bm{\delta m}(t)=\Re{\left(\left(\tilde{\delta m}_x \, \bm{e_{x}} + \tilde{\delta m}_y \, \bm{e_{y}}\right) \text{e}^{\imath(\omega t - kx)}\right)}$. The tilde recalls that the transverse terms $\tilde{\delta m}_x$ and $ \tilde{\delta m}_y$, which obey $|\tilde{\delta m}_i|\ll 1 \,(i=x,y)$, are complex numbers. 

Likewise, the total effective magnetic field $\bm{H}^{\rm eff}$ acting on $\bm{M}$ can be written as
\begin{equation}
    \bm{H}^{\rm eff}(t)=H_z^\textrm{eff} \, \bm{e_{z}} +\bm{h}^\textrm{eff}(t),
    \label{field_h}
\end{equation}
$H_z^\textrm{eff}$ is the static component of the total magnetic field. In our case, it is the sum of the out-of-plane applied field $\bm{H}_z$, the static demagnetizing field $-\bm{M}_s$, and the anisotropy field $\bm{H}_\textrm{ani}$. The dynamic magnetic field $\bm{h}^\textrm{eff}$ has two contributions: the demagnetizing field produced by the periodic magnetization pattern $- n(k) \tilde{m}_{x} \bm{e_{x}}$, with $n(k) = 1-\frac{1-e^{-\left| k\right| t}}{\left| k\right| t} \approx \frac{|k|t}{2}$ and the rf applied field $\bm{h}=\Re{\left(\left(h_x \, \bm{e_{x}} + h_y \, \bm{e_{y}}\right) \text{e}^{\imath (\omega t - kx) }\right)}$, with $h_x$ and $h_y \in \mathbb{R}$.

Keeping only the terms up to first order in the dynamic components $\bm{\delta m}$ and $\bm{h}$, the linearized LLG equation in the reciprocal space reads
\begin{equation}
    \imath\omega \bm{\delta m} = -\gamma_0 M_s(\bm{e_z} \times \bm{h}^\textrm{eff} + \bm{\delta m} \times \bm{H}_z^\textrm{eff}) + \imath\omega\alpha (\bm{e_z} \times \bm{\delta m}).
	\label{Eq_LLG}
\end{equation}

\begin{widetext}
Introducing the specific plane-wave forms chosen for $\vec{m}$ and $\vec{h}$ in Eq.~\ref{Eq_LLG} and rearranging the terms adequately, we readily obtain a matrix expression for the rf applied field as a function of the dynamic magnetization
\begin{equation}
 \gamma_0 \, M_s \begin{pmatrix}
    h_{x}\\
    h_{y}\\
    \end{pmatrix} =
    \begin{bmatrix}
    \gamma_0 (H_z^\textrm{eff} + n(k) M_s) +\imath\,\alpha\,\omega &  - \imath \, \omega\\
     \imath \, \omega & \gamma_0 \, H_z^\textrm{eff} + \imath\,\alpha\,\omega \\
    \end{bmatrix}
    \begin{pmatrix}
     \tilde{\delta m}_{x}\\
     \tilde\delta {m}_{y}\\
    \end{pmatrix}
	\label{Eq_InvsSuscep}
\end{equation}
A simple matrix inversion finally yields the susceptibility tensor $\chi$. Keeping only terms that are first order in $\alpha$, we obtain
\begin{equation}
  \begin{pmatrix}  \tilde{\delta m}_{x}\\  \tilde{\delta m}_{y}\\  \end{pmatrix} 
  = \chi \begin{pmatrix}  h_{x}\\  h_{y}\\  \end{pmatrix} 
  \approx \frac{\gamma_0 M_s}{(\omega_{k}^2 - \omega^2) + \imath \omega \Delta\omega_{k} }
  \begin{bmatrix}
    \gamma_0 \, H_z^\textrm{eff} + \imath\,\alpha\,\omega & \imath\,\omega\\
    - \imath \, \omega & \gamma_0 (H_z^\textrm{eff} + n(k) \, M_s) +\imath\,\alpha\,\omega \\
  \end{bmatrix}
  \begin{pmatrix}
    h_{x}\\
    h_{y}\\
  \end{pmatrix},
  \label{Eq_Suscep}
\end{equation}
\end{widetext}
where
\begin{equation}
    \omega_k^2=\gamma_0^2  H_z^\textrm{eff} \, (H_z^\textrm{eff} + n(k) \, M_s)
\end{equation}
is the dispersion relation of the forward volume spin waves (Eq.~\ref{Eq_dispersion}) and $\Delta \omega_k = \alpha \gamma_0 (2 \, H_z^\textrm{eff} + n(k) \, M_s)$ is the mode linewidth.

From Eq. ~\ref{Eq_Suscep}, the transverse susceptibility of interest reads
\begin{equation}
 \chi_{xx} (\omega,k) \approx \frac{\gamma_0 M_{s} ( \gamma_0 \, H_z^\textrm{eff} +\imath\,\alpha\,\omega)}{(\omega_{k}^2 - \omega^2) + \imath  \omega \Delta\omega_{k}}.
\end{equation}

Considering that $\alpha$ is much smaller than $\frac{\omega}{\gamma_0 \, H_z^\textrm{eff}} \sim 1 $, we can neglect the damping term in the numerator of the above expression, but not in the denominator since it dominates near the resonance. 

For the discussion in the main text of this work it is useful to rewrite the denominator as linear function of the wave vector $k$ for a given angular frequency $\omega_0$. At $k \sim k_0$, where $\omega_0=\omega(k_0)$, the dispersion relation can be written as $\omega_k^2=\omega^2+2 \, v_g (k_0) (k-k_0)$. Here, $v_g(k_0)$ is the group velocity at $k_0$, as defined in Eq. \ref{Eq_vg}. Thus, knowing that $\frac{\Delta \omega_k}{2\,v_g}=L_\textrm{att}^{-1}$ the susceptibility at $\omega_0$ reads 

\begin{equation}
    \chi_{xx} (\omega_0,k) \approx \frac{\gamma_0 M_{s} \gamma_0 \, H_z^\textrm{eff}}{2 \, \omega_0 \, v_g (k_0) \left(\left| k\right|-k_0+\frac{\imath}{L_\textrm{att}}\right)}.
\end{equation}

%%%%%%%%%%%%%%%%%%%%%%%%%%%%%%%%%%%%%%%%%%%%%%%%%%%
\textbf{\section*{Appendix II: Hypothetical zero damping case}}
Let's test the case with no propagation loss, i.e. assuming $\alpha=0$ and an infinite attenuation length. We assume that we possess an hypothetical narrow band antenna that carries a sinc current partition with spatial periodicity $2\pi/k_0$. The antenna thus emits uniformly in the $[k_0-\epsilon,k_0+\epsilon]$ and $[-k_0-\epsilon,-k_0+\epsilon]$ intervals and does not emit for other wavevectors. We still consider a working frequency $\omega_0 > \omega_{FMR}$ with the corresponding spin wave wavevector $k_0$, defined as positive.
The limit of the susceptibility for zero damping can be written as:
\begin{equation}
\left.\chi(\omega_0, k)\right|_{\rm \alpha \rightarrow 0} =\frac{2}{t} \frac{1}{|k|-k_0} - \imath\frac{2\pi}{t}\delta_{k_0} - \imath\frac{2\pi}{t}\delta_{-k_0}
\end{equation} where $\delta$ is the Dirac distribution.
The integration of the transmission parameter requires to split the integral in two parts:
\begin{equation}
	\tilde{T}(\omega_0) \propto \left[\int_{-k_0-\epsilon}^{-k_0+\epsilon} + \int_{k_0-\epsilon}^{k_0+\epsilon}\right] \chi(k) ~  e^{-\imath kr} dk 
	\end{equation}

After some algebra, the real part can be written as:
\begin{equation}
\left.\Re(\tilde{T}(\omega_0))\right|_{\alpha \rightarrow 0} = -\frac{4}{t} \sin ({k_0} r) \left[\pi+2 \int_ {0}^{+\epsilon r}\text {sinc} (q)\text {d} q  \right]
\label{Eq_ReTalpha0}
\end{equation}
and the imaginary part is:
\begin{equation}
\left.\Im(\tilde{T}(\omega_0))\right|_{\rm \alpha \rightarrow 0} =  -\frac{4 \pi}{t}  \cos ({k_0} r)
\end{equation}

Several points are worth noticing: \begin{itemize}
\item If the antenna had an infinitely narrow band (i.e. $\epsilon=0$, i.e. if it was generating a perfectly spatially periodic field), the transmission signal in the absence of damping would simply reduce to:
\begin{equation}
\left.\tilde{T}(\omega_0)\right|_{\alpha \rightarrow 0, \epsilon \rightarrow 0} = -\imath \frac{4 \pi}{t} e^{-\imath k_0 r}
\end{equation}
which is a trivial monochromatic plane wave propagating without attenuation, as intuitively anticipated.\\
\item If the antenna had a finite bandwidth (i.e. $\epsilon \neq 0$), the integral in Eq.~\ref{Eq_ReTalpha0} would converge to $\pi/2$ for long propagation distances, i.e. when $\epsilon r \rightarrow \infty$, such that $\left.\Re\left(\tilde{T}(\omega_0)\right|_{\alpha \rightarrow 0, \epsilon r \rightarrow \infty}\right) \rightarrow -\frac{8 \pi}{t} \sin ({k_0} r) $. The amplitude of the transmission signal would not decay with the propagation distance but it would oscillate periodically because the amplitude of the real and imaginary parts differ. This beating is due to the finite spectral width $\epsilon$ of the wavepacket.
 \\
\item Finally let's examine the case of an antenna that still has an finite spectral width (i.e. $\epsilon \neq 0$), and look at the propagation in its vicinity. Noting that $\int_{0}^{+\epsilon r} \textrm{sinc}(q)\textrm{d}q \approx \epsilon r$ for $\epsilon r \ll 1$, implies that the real part of the transmission parameter would increase with the propagation distance at small distances.
One could be surprised by this trend. This simply reflects the fact that one needs some distance so that the below-resonance wavevectors $k \in [k_0, k_0+\epsilon]$ (with positive $\chi$) and the above-resonance wavevectors $k \in [k_0-\epsilon, k_0]$  (with negative $\chi$) cease to have cancelling contributions thanks to their differential phase rotation $e^{-\imath (k-k_0) r}$ that increases with $r$.\\
\end{itemize}

\textbf{\section*{Appendix III: Effect of the spacing between the antenna and the magnetic film onto the transmission spectrum}}

\begin{figure}[htbp]
\includegraphics[width=8.6 cm]{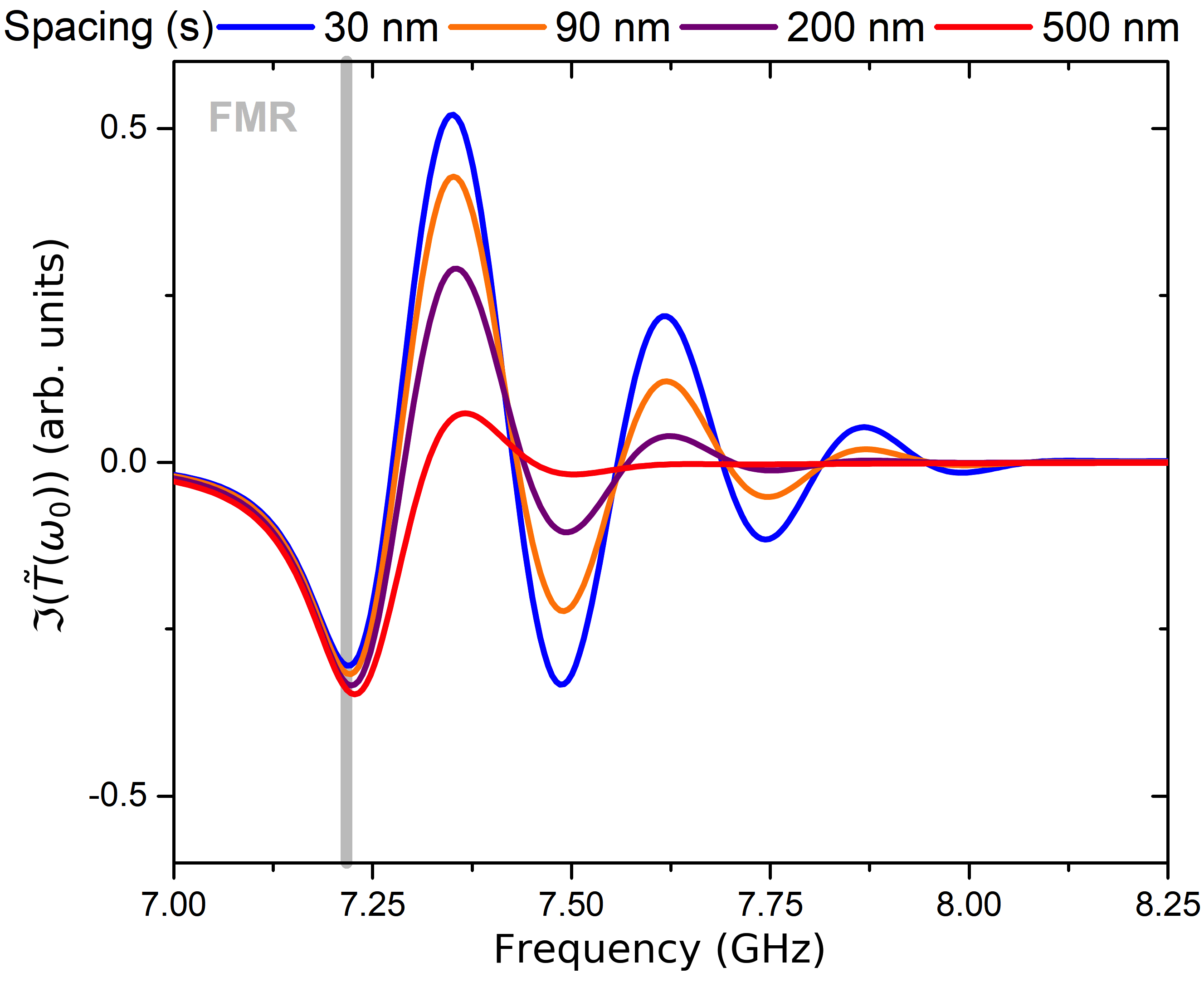}
\caption{Imaginary part of the transmission parameter for $\alpha=0.03$, CPW antenna size of $L=180$ nm and gap of $g=200$ nm and a propagation length of 1.8~$\mu\text m$. The vertical separation between the antenna and the magnetic film is varied between $s=30$ nm and $s=500$ nm.}
\label{Fig_T_vs_s}
\end{figure}

Fig.~\ref{Fig_T_vs_s} displays how the transmission parameter is expected to vary with the vertical distance $s$ between the antenna and the magnetic film. The $e^{-k s}$ factor in Eq.~\ref{Eq_AeffS} does not change the response at the FMR frequency but it induces a faster decay of the transmission signal versus frequency above FMR, and a progressive loss of the oscillatory character. As a result, $s$ and $\alpha$ have qualitatively similar influences on the lineshape of the transmission signal, although $\alpha$ exponentially impacts the amplitude of the transmission. Note that in our electromagnetic model (Eqs.~\ref{Eq_AeffS}-\ref{Eq_AeffGSG}), the antenna and the film are approximated as infinitely thin sheets, such that there is no rigorous way of choosing the right value of $s$ to fit the experiment.

%%%%%%%%%%%%%%%%%%%%%%%%%%%%%%%%%%%%%%%%%%%%%%%%%%%
%%%%%%%%%%%%%%%%%%%%%%%%%%%%%%%%%%%%%%%%%%%%%%%%%%%

\bibliography{References}

\begin{thebibliography}{28}%
\makeatletter
\providecommand \@ifxundefined [1]{%
 \@ifx{#1\undefined}
}%
\providecommand \@ifnum [1]{%
 \ifnum #1\expandafter \@firstoftwo
 \else \expandafter \@secondoftwo
 \fi
}%
\providecommand \@ifx [1]{%
 \ifx #1\expandafter \@firstoftwo
 \else \expandafter \@secondoftwo
 \fi
}%
\providecommand \natexlab [1]{#1}%
\providecommand \enquote  [1]{``#1''}%
\providecommand \bibnamefont  [1]{#1}%
\providecommand \bibfnamefont [1]{#1}%
\providecommand \citenamefont [1]{#1}%
\providecommand \href@noop [0]{\@secondoftwo}%
\providecommand \href [0]{\begingroup \@sanitize@url \@href}%
\providecommand \@href[1]{\@@startlink{#1}\@@href}%
\providecommand \@@href[1]{\endgroup#1\@@endlink}%
\providecommand \@sanitize@url [0]{\catcode `\\12\catcode `\$12\catcode
  `\&12\catcode `\#12\catcode `\^12\catcode `\_12\catcode `\%12\relax}%
\providecommand \@@startlink[1]{}%
\providecommand \@@endlink[0]{}%
\providecommand \url  [0]{\begingroup\@sanitize@url \@url }%
\providecommand \@url [1]{\endgroup\@href {#1}{\urlprefix }}%
\providecommand \urlprefix  [0]{URL }%
\providecommand \Eprint [0]{\href }%
\providecommand \doibase [0]{http://dx.doi.org/}%
\providecommand \selectlanguage [0]{\@gobble}%
\providecommand \bibinfo  [0]{\@secondoftwo}%
\providecommand \bibfield  [0]{\@secondoftwo}%
\providecommand \translation [1]{[#1]}%
\providecommand \BibitemOpen [0]{}%
\providecommand \bibitemStop [0]{}%
\providecommand \bibitemNoStop [0]{.\EOS\space}%
\providecommand \EOS [0]{\spacefactor3000\relax}%
\providecommand \BibitemShut  [1]{\csname bibitem#1\endcsname}%
\let\auto@bib@innerbib\@empty
%</preamble>
\bibitem [{\citenamefont {Chumak}\ \emph {et~al.}(2015)\citenamefont {Chumak},
  \citenamefont {Vasyuchka}, \citenamefont {Serga},\ and\ \citenamefont
  {Hillebrands}}]{Chumak2015}%
  \BibitemOpen
  \bibfield  {author} {\bibinfo {author} {\bibfnamefont {A.~V.}\ \bibnamefont
  {Chumak}}, \bibinfo {author} {\bibfnamefont {V.~I.}\ \bibnamefont
  {Vasyuchka}}, \bibinfo {author} {\bibfnamefont {A.~A.}\ \bibnamefont
  {Serga}}, \ and\ \bibinfo {author} {\bibfnamefont {B.}~\bibnamefont
  {Hillebrands}},\ }\href {\doibase 10.1038/nphys3347} {\bibfield  {journal}
  {\bibinfo  {journal} {Nature Physics}\ }\textbf {\bibinfo {volume} {11}},\
  \bibinfo {pages} {453} (\bibinfo {year} {2015})}\BibitemShut {NoStop}%
\bibitem [{\citenamefont {Vlaminck}\ and\ \citenamefont
  {Bailleul}(2008)}]{Vlaminck2008}%
  \BibitemOpen
  \bibfield  {author} {\bibinfo {author} {\bibfnamefont {V.}~\bibnamefont
  {Vlaminck}}\ and\ \bibinfo {author} {\bibfnamefont {M.}~\bibnamefont
  {Bailleul}},\ }\href {\doibase 10.1126/science.1162843} {\bibfield  {journal}
  {\bibinfo  {journal} {Science}\ }\textbf {\bibinfo {volume} {322}},\ \bibinfo
  {pages} {410} (\bibinfo {year} {2008})}\BibitemShut {NoStop}%
\bibitem [{\citenamefont {Wagner}\ \emph {et~al.}(2016)\citenamefont {Wagner},
  \citenamefont {K{\'{a}}kay}, \citenamefont {Schultheiss}, \citenamefont
  {Henschke}, \citenamefont {Sebastian},\ and\ \citenamefont
  {Schultheiss}}]{Wagner2016}%
  \BibitemOpen
  \bibfield  {author} {\bibinfo {author} {\bibfnamefont {K.}~\bibnamefont
  {Wagner}}, \bibinfo {author} {\bibfnamefont {A.}~\bibnamefont {K{\'{a}}kay}},
  \bibinfo {author} {\bibfnamefont {K.}~\bibnamefont {Schultheiss}}, \bibinfo
  {author} {\bibfnamefont {A.}~\bibnamefont {Henschke}}, \bibinfo {author}
  {\bibfnamefont {T.}~\bibnamefont {Sebastian}}, \ and\ \bibinfo {author}
  {\bibfnamefont {H.}~\bibnamefont {Schultheiss}},\ }\href {\doibase
  10.1038/nnano.2015.339} {\bibfield  {journal} {\bibinfo  {journal} {Nature
  Nanotechnology}\ }\textbf {\bibinfo {volume} {11}},\ \bibinfo {pages} {432}
  (\bibinfo {year} {2016})}\BibitemShut {NoStop}%
\bibitem [{\citenamefont {Vogt}\ \emph {et~al.}(2014)\citenamefont {Vogt},
  \citenamefont {Fradin}, \citenamefont {Pearson}, \citenamefont {Sebastian},
  \citenamefont {Bader}, \citenamefont {Hillebrands}, \citenamefont
  {Hoffmann},\ and\ \citenamefont {Schultheiss}}]{Vogt2014}%
  \BibitemOpen
  \bibfield  {author} {\bibinfo {author} {\bibfnamefont {K.}~\bibnamefont
  {Vogt}}, \bibinfo {author} {\bibfnamefont {F.}~\bibnamefont {Fradin}},
  \bibinfo {author} {\bibfnamefont {J.}~\bibnamefont {Pearson}}, \bibinfo
  {author} {\bibfnamefont {T.}~\bibnamefont {Sebastian}}, \bibinfo {author}
  {\bibfnamefont {S.}~\bibnamefont {Bader}}, \bibinfo {author} {\bibfnamefont
  {B.}~\bibnamefont {Hillebrands}}, \bibinfo {author} {\bibfnamefont
  {A.}~\bibnamefont {Hoffmann}}, \ and\ \bibinfo {author} {\bibfnamefont
  {H.}~\bibnamefont {Schultheiss}},\ }\href {\doibase 10.1038/ncomms4727}
  {\bibfield  {journal} {\bibinfo  {journal} {Nature Communications}\ }\textbf
  {\bibinfo {volume} {5}} (\bibinfo {year} {2014}),\
  10.1038/ncomms4727}\BibitemShut {NoStop}%
\bibitem [{\citenamefont {Khitun}\ \emph {et~al.}(2010)\citenamefont {Khitun},
  \citenamefont {Bao},\ and\ \citenamefont {Wang}}]{Khitun2010}%
  \BibitemOpen
  \bibfield  {author} {\bibinfo {author} {\bibfnamefont {A.}~\bibnamefont
  {Khitun}}, \bibinfo {author} {\bibfnamefont {M.}~\bibnamefont {Bao}}, \ and\
  \bibinfo {author} {\bibfnamefont {K.~L.}\ \bibnamefont {Wang}},\ }\href
  {\doibase 10.1088/0022-3727/43/26/264005} {\bibfield  {journal} {\bibinfo
  {journal} {Journal of Physics D: Applied Physics}\ }\textbf {\bibinfo
  {volume} {43}},\ \bibinfo {pages} {264005} (\bibinfo {year}
  {2010})}\BibitemShut {NoStop}%
\bibitem [{\citenamefont {Klingler}\ \emph {et~al.}(2015)\citenamefont
  {Klingler}, \citenamefont {Pirro}, \citenamefont {Brächer}, \citenamefont
  {Leven}, \citenamefont {Hillebrands},\ and\ \citenamefont
  {Chumak}}]{Klingler2015}%
  \BibitemOpen
  \bibfield  {author} {\bibinfo {author} {\bibfnamefont {S.}~\bibnamefont
  {Klingler}}, \bibinfo {author} {\bibfnamefont {P.}~\bibnamefont {Pirro}},
  \bibinfo {author} {\bibfnamefont {T.}~\bibnamefont {Brächer}}, \bibinfo
  {author} {\bibfnamefont {B.}~\bibnamefont {Leven}}, \bibinfo {author}
  {\bibfnamefont {B.}~\bibnamefont {Hillebrands}}, \ and\ \bibinfo {author}
  {\bibfnamefont {A.~V.}\ \bibnamefont {Chumak}},\ }\href {\doibase
  10.1063/1.4921850} {\bibfield  {journal} {\bibinfo  {journal} {Applied
  Physics Letters}\ }\textbf {\bibinfo {volume} {106}},\ \bibinfo {pages}
  {212406} (\bibinfo {year} {2015})}\BibitemShut {NoStop}%
\bibitem [{\citenamefont {Grundler}(2015)}]{Grundler2015}%
  \BibitemOpen
  \bibfield  {author} {\bibinfo {author} {\bibfnamefont {D.}~\bibnamefont
  {Grundler}},\ }\href {\doibase 10.1038/nphys3349} {\bibfield  {journal}
  {\bibinfo  {journal} {Nature Physics}\ }\textbf {\bibinfo {volume} {11}},\
  \bibinfo {pages} {438} (\bibinfo {year} {2015})}\BibitemShut {NoStop}%
\bibitem [{\citenamefont {Schneider}\ \emph {et~al.}(2008)\citenamefont
  {Schneider}, \citenamefont {Serga}, \citenamefont {Leven}, \citenamefont
  {Hillebrands}, \citenamefont {Stamps},\ and\ \citenamefont
  {Kostylev}}]{Schneider2008}%
  \BibitemOpen
  \bibfield  {author} {\bibinfo {author} {\bibfnamefont {T.}~\bibnamefont
  {Schneider}}, \bibinfo {author} {\bibfnamefont {A.~A.}\ \bibnamefont
  {Serga}}, \bibinfo {author} {\bibfnamefont {B.}~\bibnamefont {Leven}},
  \bibinfo {author} {\bibfnamefont {B.}~\bibnamefont {Hillebrands}}, \bibinfo
  {author} {\bibfnamefont {R.~L.}\ \bibnamefont {Stamps}}, \ and\ \bibinfo
  {author} {\bibfnamefont {M.~P.}\ \bibnamefont {Kostylev}},\ }\href {\doibase
  10.1063/1.2834714} {\bibfield  {journal} {\bibinfo  {journal} {Applied
  Physics Letters}\ }\textbf {\bibinfo {volume} {92}},\ \bibinfo {pages}
  {022505} (\bibinfo {year} {2008})}\BibitemShut {NoStop}%
\bibitem [{\citenamefont {Demidov}\ and\ \citenamefont
  {Demokritov}(2015)}]{Demidov2015}%
  \BibitemOpen
  \bibfield  {author} {\bibinfo {author} {\bibfnamefont {V.~E.}\ \bibnamefont
  {Demidov}}\ and\ \bibinfo {author} {\bibfnamefont {S.~O.}\ \bibnamefont
  {Demokritov}},\ }\href {\doibase 10.1109/tmag.2014.2388196} {\bibfield
  {journal} {\bibinfo  {journal} {{IEEE} Transactions on Magnetics}\ }\textbf
  {\bibinfo {volume} {51}},\ \bibinfo {pages} {1} (\bibinfo {year}
  {2015})}\BibitemShut {NoStop}%
\bibitem [{\citenamefont {Talmelli}\ \emph {et~al.}(2018)\citenamefont
  {Talmelli}, \citenamefont {Ciubotaru}, \citenamefont {Garello}, \citenamefont
  {Sun}, \citenamefont {Heyns}, \citenamefont {Radu}, \citenamefont
  {Adelmann},\ and\ \citenamefont {Devolder}}]{Talmelli2018}%
  \BibitemOpen
  \bibfield  {author} {\bibinfo {author} {\bibfnamefont {G.}~\bibnamefont
  {Talmelli}}, \bibinfo {author} {\bibfnamefont {F.}~\bibnamefont {Ciubotaru}},
  \bibinfo {author} {\bibfnamefont {K.}~\bibnamefont {Garello}}, \bibinfo
  {author} {\bibfnamefont {X.}~\bibnamefont {Sun}}, \bibinfo {author}
  {\bibfnamefont {M.}~\bibnamefont {Heyns}}, \bibinfo {author} {\bibfnamefont
  {I.~P.}\ \bibnamefont {Radu}}, \bibinfo {author} {\bibfnamefont
  {C.}~\bibnamefont {Adelmann}}, \ and\ \bibinfo {author} {\bibfnamefont
  {T.}~\bibnamefont {Devolder}},\ }\href {\doibase
  10.1103/physrevapplied.10.044060} {\bibfield  {journal} {\bibinfo  {journal}
  {Physical Review Applied}\ }\textbf {\bibinfo {volume} {10}} (\bibinfo {year}
  {2018}),\ 10.1103/physrevapplied.10.044060}\BibitemShut {NoStop}%
\bibitem [{\citenamefont {Haldar}\ \emph {et~al.}(2016)\citenamefont {Haldar},
  \citenamefont {Kumar},\ and\ \citenamefont {Adeyeye}}]{Haldar2016}%
  \BibitemOpen
  \bibfield  {author} {\bibinfo {author} {\bibfnamefont {A.}~\bibnamefont
  {Haldar}}, \bibinfo {author} {\bibfnamefont {D.}~\bibnamefont {Kumar}}, \
  and\ \bibinfo {author} {\bibfnamefont {A.~O.}\ \bibnamefont {Adeyeye}},\
  }\href {\doibase 10.1038/nnano.2015.332} {\bibfield  {journal} {\bibinfo
  {journal} {Nature Nanotechnology}\ }\textbf {\bibinfo {volume} {11}},\
  \bibinfo {pages} {437} (\bibinfo {year} {2016})}\BibitemShut {NoStop}%
\bibitem [{\citenamefont {Bhaskar}\ \emph {et~al.}(2020)\citenamefont
  {Bhaskar}, \citenamefont {Talmelli}, \citenamefont {Ciubotaru}, \citenamefont
  {Adelmann},\ and\ \citenamefont {Devolder}}]{bhaskar_backward_2020}%
  \BibitemOpen
  \bibfield  {author} {\bibinfo {author} {\bibfnamefont {U.~K.}\ \bibnamefont
  {Bhaskar}}, \bibinfo {author} {\bibfnamefont {G.}~\bibnamefont {Talmelli}},
  \bibinfo {author} {\bibfnamefont {F.}~\bibnamefont {Ciubotaru}}, \bibinfo
  {author} {\bibfnamefont {C.}~\bibnamefont {Adelmann}}, \ and\ \bibinfo
  {author} {\bibfnamefont {T.}~\bibnamefont {Devolder}},\ }\href {\doibase
  10.1063/1.5125751} {\bibfield  {journal} {\bibinfo  {journal} {Journal of
  Applied Physics}\ }\textbf {\bibinfo {volume} {127}},\ \bibinfo {pages}
  {033902} (\bibinfo {year} {2020})},\ \Eprint
  {http://arxiv.org/abs/https://doi.org/10.1063/1.5125751}
  {https://doi.org/10.1063/1.5125751} \BibitemShut {NoStop}%
\bibitem [{\citenamefont {Kalinikos}\ and\ \citenamefont
  {Slavin}(1986)}]{Kalinikos1986}%
  \BibitemOpen
  \bibfield  {author} {\bibinfo {author} {\bibfnamefont {B.~A.}\ \bibnamefont
  {Kalinikos}}\ and\ \bibinfo {author} {\bibfnamefont {A.~N.}\ \bibnamefont
  {Slavin}},\ }\href {\doibase 10.1088/0022-3719/19/35/014} {\bibfield
  {journal} {\bibinfo  {journal} {Journal of Physics C: Solid State Physics}\
  }\textbf {\bibinfo {volume} {19}},\ \bibinfo {pages} {7013} (\bibinfo {year}
  {1986})}\BibitemShut {NoStop}%
\bibitem [{\citenamefont {Kanazawa}\ \emph {et~al.}(2016)\citenamefont
  {Kanazawa}, \citenamefont {Goto}, \citenamefont {Sekiguchi}, \citenamefont
  {Granovsky}, \citenamefont {Ross}, \citenamefont {Takagi}, \citenamefont
  {Nakamura},\ and\ \citenamefont {Inoue}}]{Kanazawa2016}%
  \BibitemOpen
  \bibfield  {author} {\bibinfo {author} {\bibfnamefont {N.}~\bibnamefont
  {Kanazawa}}, \bibinfo {author} {\bibfnamefont {T.}~\bibnamefont {Goto}},
  \bibinfo {author} {\bibfnamefont {K.}~\bibnamefont {Sekiguchi}}, \bibinfo
  {author} {\bibfnamefont {A.~B.}\ \bibnamefont {Granovsky}}, \bibinfo {author}
  {\bibfnamefont {C.~A.}\ \bibnamefont {Ross}}, \bibinfo {author}
  {\bibfnamefont {H.}~\bibnamefont {Takagi}}, \bibinfo {author} {\bibfnamefont
  {Y.}~\bibnamefont {Nakamura}}, \ and\ \bibinfo {author} {\bibfnamefont
  {M.}~\bibnamefont {Inoue}},\ }\href {\doibase 10.1038/srep30268} {\bibfield
  {journal} {\bibinfo  {journal} {Scientific Reports}\ }\textbf {\bibinfo
  {volume} {6}} (\bibinfo {year} {2016}),\ 10.1038/srep30268}\BibitemShut
  {NoStop}%
\bibitem [{\citenamefont {Chen}\ \emph {et~al.}(2018)\citenamefont {Chen},
  \citenamefont {F.}, \citenamefont {Liu}, \citenamefont {Yu}, \citenamefont
  {Liu}, \citenamefont {Chang}, \citenamefont {Stöckler}, \citenamefont {Hu},
  \citenamefont {Zeng}, \citenamefont {Zhang}, \citenamefont {Liao},
  \citenamefont {Yu}, \citenamefont {Zhao},\ and\ \citenamefont
  {Wu}}]{Chen2018}%
  \BibitemOpen
  \bibfield  {author} {\bibinfo {author} {\bibfnamefont {J.}~\bibnamefont
  {Chen}}, \bibinfo {author} {\bibfnamefont {H.}~\bibnamefont {F.}}, \bibinfo
  {author} {\bibfnamefont {T.}~\bibnamefont {Liu}}, \bibinfo {author}
  {\bibfnamefont {H.}~\bibnamefont {Yu}}, \bibinfo {author} {\bibfnamefont
  {C.}~\bibnamefont {Liu}}, \bibinfo {author} {\bibfnamefont {H.}~\bibnamefont
  {Chang}}, \bibinfo {author} {\bibfnamefont {T.}~\bibnamefont {Stöckler}},
  \bibinfo {author} {\bibfnamefont {J.}~\bibnamefont {Hu}}, \bibinfo {author}
  {\bibfnamefont {L.}~\bibnamefont {Zeng}}, \bibinfo {author} {\bibfnamefont
  {Y.}~\bibnamefont {Zhang}}, \bibinfo {author} {\bibfnamefont
  {Z.}~\bibnamefont {Liao}}, \bibinfo {author} {\bibfnamefont {D.}~\bibnamefont
  {Yu}}, \bibinfo {author} {\bibfnamefont {W.}~\bibnamefont {Zhao}}, \ and\
  \bibinfo {author} {\bibfnamefont {M.}~\bibnamefont {Wu}},\ }\href {\doibase
  https://doi.org/10.1016/j.jmmm.2017.04.045} {\bibfield  {journal} {\bibinfo
  {journal} {J. Magn. Magn. Mater.}\ }\textbf {\bibinfo {volume} {450}},\
  \bibinfo {pages} {3} (\bibinfo {year} {2018})}\BibitemShut {NoStop}%
\bibitem [{\citenamefont {Han}\ \emph {et~al.}(2019)\citenamefont {Han},
  \citenamefont {Zhang}, \citenamefont {Hou}, \citenamefont {Siddiqui},\ and\
  \citenamefont {Liu}}]{Han2019}%
  \BibitemOpen
  \bibfield  {author} {\bibinfo {author} {\bibfnamefont {J.}~\bibnamefont
  {Han}}, \bibinfo {author} {\bibfnamefont {P.}~\bibnamefont {Zhang}}, \bibinfo
  {author} {\bibfnamefont {J.~T.}\ \bibnamefont {Hou}}, \bibinfo {author}
  {\bibfnamefont {S.~A.}\ \bibnamefont {Siddiqui}}, \ and\ \bibinfo {author}
  {\bibfnamefont {L.}~\bibnamefont {Liu}},\ }\href {\doibase
  10.1126/science.aau2610} {\bibfield  {journal} {\bibinfo  {journal}
  {Science}\ }\textbf {\bibinfo {volume} {366}},\ \bibinfo {pages} {1121}
  (\bibinfo {year} {2019})}\BibitemShut {NoStop}%
\bibitem [{\citenamefont {Vlaminck}\ and\ \citenamefont
  {Bailleul}(2010)}]{Vlaminck2010}%
  \BibitemOpen
  \bibfield  {author} {\bibinfo {author} {\bibfnamefont {V.}~\bibnamefont
  {Vlaminck}}\ and\ \bibinfo {author} {\bibfnamefont {M.}~\bibnamefont
  {Bailleul}},\ }\href {\doibase 10.1103/physrevb.81.014425} {\bibfield
  {journal} {\bibinfo  {journal} {Physical Review B}\ }\textbf {\bibinfo
  {volume} {81}} (\bibinfo {year} {2010}),\
  10.1103/physrevb.81.014425}\BibitemShut {NoStop}%
\bibitem [{\citenamefont {Gladii}\ \emph {et~al.}(2016)\citenamefont {Gladii},
  \citenamefont {Haidar}, \citenamefont {Henry}, \citenamefont {Kostylev},\
  and\ \citenamefont {Bailleul}}]{Gladii2016}%
  \BibitemOpen
  \bibfield  {author} {\bibinfo {author} {\bibfnamefont {O.}~\bibnamefont
  {Gladii}}, \bibinfo {author} {\bibfnamefont {M.}~\bibnamefont {Haidar}},
  \bibinfo {author} {\bibfnamefont {Y.}~\bibnamefont {Henry}}, \bibinfo
  {author} {\bibfnamefont {M.}~\bibnamefont {Kostylev}}, \ and\ \bibinfo
  {author} {\bibfnamefont {M.}~\bibnamefont {Bailleul}},\ }\href {\doibase
  10.1103/physrevb.93.054430} {\bibfield  {journal} {\bibinfo  {journal}
  {Physical Review B}\ }\textbf {\bibinfo {volume} {93}} (\bibinfo {year}
  {2016}),\ 10.1103/physrevb.93.054430}\BibitemShut {NoStop}%
\bibitem [{\citenamefont {Stancil}\ and\ \citenamefont
  {Prabhakar}(2009)}]{2009a}%
  \BibitemOpen
  \bibfield  {author} {\bibinfo {author} {\bibfnamefont {D.}~\bibnamefont
  {Stancil}}\ and\ \bibinfo {author} {\bibfnamefont {A.}~\bibnamefont
  {Prabhakar}},\ }\href {\doibase 10.1007/978-0-387-77865-5} {\emph {\bibinfo
  {title} {Spin Waves: Theory and Applications}}}\ (\bibinfo  {publisher}
  {Springer {US}},\ \bibinfo {year} {2009})\BibitemShut {NoStop}%
\bibitem [{\citenamefont {Ciubotaru}\ \emph {et~al.}(2016)\citenamefont
  {Ciubotaru}, \citenamefont {Devolder}, \citenamefont {Manfrini},
  \citenamefont {Adelmann},\ and\ \citenamefont {Radu}}]{Ciubotaru2016}%
  \BibitemOpen
  \bibfield  {author} {\bibinfo {author} {\bibfnamefont {F.}~\bibnamefont
  {Ciubotaru}}, \bibinfo {author} {\bibfnamefont {T.}~\bibnamefont {Devolder}},
  \bibinfo {author} {\bibfnamefont {M.}~\bibnamefont {Manfrini}}, \bibinfo
  {author} {\bibfnamefont {C.}~\bibnamefont {Adelmann}}, \ and\ \bibinfo
  {author} {\bibfnamefont {I.~P.}\ \bibnamefont {Radu}},\ }\href {\doibase
  10.1063/1.4955030} {\bibfield  {journal} {\bibinfo  {journal} {Applied
  Physics Letters}\ }\textbf {\bibinfo {volume} {109}},\ \bibinfo {pages}
  {012403} (\bibinfo {year} {2016})}\BibitemShut {NoStop}%
\bibitem [{\citenamefont {Talmelli}\ \emph {et~al.}()\citenamefont {Talmelli},
  \citenamefont {Devolder}, \citenamefont {Träger}, \citenamefont {Förster},
  \citenamefont {Wintz}, \citenamefont {Weigand}, \citenamefont {Stoll},
  \citenamefont {Heyns}, \citenamefont {Schütz}, \citenamefont {Radu},
  \citenamefont {Gräfe}, \citenamefont {Ciubotaru},\ and\ \citenamefont
  {Adelmann}}]{Talmelli2019}%
  \BibitemOpen
  \bibfield  {author} {\bibinfo {author} {\bibfnamefont {G.}~\bibnamefont
  {Talmelli}}, \bibinfo {author} {\bibfnamefont {T.}~\bibnamefont {Devolder}},
  \bibinfo {author} {\bibfnamefont {N.}~\bibnamefont {Träger}}, \bibinfo
  {author} {\bibfnamefont {J.}~\bibnamefont {Förster}}, \bibinfo {author}
  {\bibfnamefont {S.}~\bibnamefont {Wintz}}, \bibinfo {author} {\bibfnamefont
  {M.}~\bibnamefont {Weigand}}, \bibinfo {author} {\bibfnamefont
  {H.}~\bibnamefont {Stoll}}, \bibinfo {author} {\bibfnamefont
  {M.}~\bibnamefont {Heyns}}, \bibinfo {author} {\bibfnamefont
  {G.}~\bibnamefont {Schütz}}, \bibinfo {author} {\bibfnamefont
  {I.}~\bibnamefont {Radu}}, \bibinfo {author} {\bibfnamefont {J.}~\bibnamefont
  {Gräfe}}, \bibinfo {author} {\bibfnamefont {F.}~\bibnamefont {Ciubotaru}}, \
  and\ \bibinfo {author} {\bibfnamefont {C.}~\bibnamefont {Adelmann}},\
  }\href@noop {} {\ }\Eprint
  {http://arxiv.org/abs/http://arxiv.org/abs/1908.02546v2}
  {http://arxiv.org/abs/1908.02546v2} \BibitemShut {NoStop}%
\bibitem [{\citenamefont {Devolder}\ \emph {et~al.}(2013)\citenamefont
  {Devolder}, \citenamefont {Ducrot}, \citenamefont {Adam}, \citenamefont
  {Barisic}, \citenamefont {Vernier}, \citenamefont {Kim}, \citenamefont
  {Ockert},\ and\ \citenamefont {Ravelosona}}]{Devolder2013}%
  \BibitemOpen
  \bibfield  {author} {\bibinfo {author} {\bibfnamefont {T.}~\bibnamefont
  {Devolder}}, \bibinfo {author} {\bibfnamefont {P.-H.}\ \bibnamefont
  {Ducrot}}, \bibinfo {author} {\bibfnamefont {J.-P.}\ \bibnamefont {Adam}},
  \bibinfo {author} {\bibfnamefont {I.}~\bibnamefont {Barisic}}, \bibinfo
  {author} {\bibfnamefont {N.}~\bibnamefont {Vernier}}, \bibinfo {author}
  {\bibfnamefont {J.-V.}\ \bibnamefont {Kim}}, \bibinfo {author} {\bibfnamefont
  {B.}~\bibnamefont {Ockert}}, \ and\ \bibinfo {author} {\bibfnamefont
  {D.}~\bibnamefont {Ravelosona}},\ }\href {\doibase 10.1063/1.4775684}
  {\bibfield  {journal} {\bibinfo  {journal} {Applied Physics Letters}\
  }\textbf {\bibinfo {volume} {102}},\ \bibinfo {pages} {022407} (\bibinfo
  {year} {2013})}\BibitemShut {NoStop}%
\bibitem [{\citenamefont {Kostylev}\ \emph {et~al.}(2007)\citenamefont
  {Kostylev}, \citenamefont {Serga}, \citenamefont {Schneider}, \citenamefont
  {Neumann}, \citenamefont {Leven}, \citenamefont {Hillebrands},\ and\
  \citenamefont {Stamps}}]{Kostylev2007}%
  \BibitemOpen
  \bibfield  {author} {\bibinfo {author} {\bibfnamefont {M.~P.}\ \bibnamefont
  {Kostylev}}, \bibinfo {author} {\bibfnamefont {A.~A.}\ \bibnamefont {Serga}},
  \bibinfo {author} {\bibfnamefont {T.}~\bibnamefont {Schneider}}, \bibinfo
  {author} {\bibfnamefont {T.}~\bibnamefont {Neumann}}, \bibinfo {author}
  {\bibfnamefont {B.}~\bibnamefont {Leven}}, \bibinfo {author} {\bibfnamefont
  {B.}~\bibnamefont {Hillebrands}}, \ and\ \bibinfo {author} {\bibfnamefont
  {R.~L.}\ \bibnamefont {Stamps}},\ }\href {\doibase
  10.1103/PhysRevB.76.184419} {\bibfield  {journal} {\bibinfo  {journal} {Phys.
  Rev. B}\ }\textbf {\bibinfo {volume} {76}},\ \bibinfo {pages} {184419}
  (\bibinfo {year} {2007})}\BibitemShut {NoStop}%
\bibitem [{\citenamefont {Collet}\ \emph {et~al.}(2017)\citenamefont {Collet},
  \citenamefont {Gladii}, \citenamefont {Evelt}, \citenamefont {Bessonov},
  \citenamefont {Soumah}, \citenamefont {Bortolotti}, \citenamefont
  {Demokritov}, \citenamefont {Henry}, \citenamefont {Cros}, \citenamefont
  {Bailleul}, \citenamefont {Demidov},\ and\ \citenamefont
  {Anane}}]{Collet2017}%
  \BibitemOpen
  \bibfield  {author} {\bibinfo {author} {\bibfnamefont {M.}~\bibnamefont
  {Collet}}, \bibinfo {author} {\bibfnamefont {O.}~\bibnamefont {Gladii}},
  \bibinfo {author} {\bibfnamefont {M.}~\bibnamefont {Evelt}}, \bibinfo
  {author} {\bibfnamefont {V.}~\bibnamefont {Bessonov}}, \bibinfo {author}
  {\bibfnamefont {L.}~\bibnamefont {Soumah}}, \bibinfo {author} {\bibfnamefont
  {P.}~\bibnamefont {Bortolotti}}, \bibinfo {author} {\bibfnamefont {S.~O.}\
  \bibnamefont {Demokritov}}, \bibinfo {author} {\bibfnamefont
  {Y.}~\bibnamefont {Henry}}, \bibinfo {author} {\bibfnamefont
  {V.}~\bibnamefont {Cros}}, \bibinfo {author} {\bibfnamefont {M.}~\bibnamefont
  {Bailleul}}, \bibinfo {author} {\bibfnamefont {V.~E.}\ \bibnamefont
  {Demidov}}, \ and\ \bibinfo {author} {\bibfnamefont {A.}~\bibnamefont
  {Anane}},\ }\href {\doibase 10.1063/1.4976708} {\bibfield  {journal}
  {\bibinfo  {journal} {Applied Physics Letters}\ }\textbf {\bibinfo {volume}
  {110}},\ \bibinfo {pages} {092408} (\bibinfo {year} {2017})}\BibitemShut
  {NoStop}%
\bibitem [{\citenamefont {Pirro}\ \emph {et~al.}(2014)\citenamefont {Pirro},
  \citenamefont {Bracher}, \citenamefont {Chumak}, \citenamefont {Lagel},
  \citenamefont {Dubs}, \citenamefont {Surzhenko}, \citenamefont {Gornert},
  \citenamefont {Leven},\ and\ \citenamefont {Hillebrands}}]{Pirro2014}%
  \BibitemOpen
  \bibfield  {author} {\bibinfo {author} {\bibfnamefont {P.}~\bibnamefont
  {Pirro}}, \bibinfo {author} {\bibfnamefont {T.}~\bibnamefont {Bracher}},
  \bibinfo {author} {\bibfnamefont {A.~V.}\ \bibnamefont {Chumak}}, \bibinfo
  {author} {\bibfnamefont {B.}~\bibnamefont {Lagel}}, \bibinfo {author}
  {\bibfnamefont {C.}~\bibnamefont {Dubs}}, \bibinfo {author} {\bibfnamefont
  {O.}~\bibnamefont {Surzhenko}}, \bibinfo {author} {\bibfnamefont
  {P.}~\bibnamefont {Gornert}}, \bibinfo {author} {\bibfnamefont
  {B.}~\bibnamefont {Leven}}, \ and\ \bibinfo {author} {\bibfnamefont
  {B.}~\bibnamefont {Hillebrands}},\ }\href {\doibase 10.1063/1.4861343}
  {\bibfield  {journal} {\bibinfo  {journal} {Applied Physics Letters}\
  }\textbf {\bibinfo {volume} {104}},\ \bibinfo {pages} {012402} (\bibinfo
  {year} {2014})}\BibitemShut {NoStop}%
\bibitem [{Note1()}]{Note1}%
  \BibitemOpen
  \bibinfo {note} {In a corpuscular approach, $L_\protect \textrm {att}$
  describes how an hypothetical 'pure' population of identical particles
  (magnons) sharing the same unique wave vector and the same unique frequency
  would progressively disappear upon propagation by loosing energy to the
  \protect \textit {non-magnetic} degrees of freedom. However, this way of
  apprehending the facts is quite hypothetical since getting a 'pure'
  population of identical magnons would require a perfectly periodic antenna
  extending to both infinities, a situation where the concept of spatial
  attenuation cannot be defined.}\BibitemShut {Stop}%
\bibitem [{\citenamefont {Birt}\ \emph {et~al.}(2012)\citenamefont {Birt},
  \citenamefont {An}, \citenamefont {Tsoi}, \citenamefont {Tamaru},
  \citenamefont {Ricketts}, \citenamefont {Wong}, \citenamefont {Amiri},
  \citenamefont {Wang},\ and\ \citenamefont {Li}}]{Birt2012}%
  \BibitemOpen
  \bibfield  {author} {\bibinfo {author} {\bibfnamefont {D.~R.}\ \bibnamefont
  {Birt}}, \bibinfo {author} {\bibfnamefont {K.}~\bibnamefont {An}}, \bibinfo
  {author} {\bibfnamefont {M.}~\bibnamefont {Tsoi}}, \bibinfo {author}
  {\bibfnamefont {S.}~\bibnamefont {Tamaru}}, \bibinfo {author} {\bibfnamefont
  {D.}~\bibnamefont {Ricketts}}, \bibinfo {author} {\bibfnamefont {K.~L.}\
  \bibnamefont {Wong}}, \bibinfo {author} {\bibfnamefont {P.~K.}\ \bibnamefont
  {Amiri}}, \bibinfo {author} {\bibfnamefont {K.~L.}\ \bibnamefont {Wang}}, \
  and\ \bibinfo {author} {\bibfnamefont {X.}~\bibnamefont {Li}},\ }\href
  {\doibase 10.1063/1.4772798} {\bibfield  {journal} {\bibinfo  {journal}
  {Applied Physics Letters}\ }\textbf {\bibinfo {volume} {101}},\ \bibinfo
  {pages} {252409} (\bibinfo {year} {2012})}\BibitemShut {NoStop}%
\bibitem [{\citenamefont {Demidov}\ \emph {et~al.}(2016)\citenamefont
  {Demidov}, \citenamefont {Urazhdin}, \citenamefont {Liu}, \citenamefont
  {Divinskiy}, \citenamefont {Telegin},\ and\ \citenamefont
  {Demokritov}}]{Demidov2016}%
  \BibitemOpen
  \bibfield  {author} {\bibinfo {author} {\bibfnamefont {V.~E.}\ \bibnamefont
  {Demidov}}, \bibinfo {author} {\bibfnamefont {S.}~\bibnamefont {Urazhdin}},
  \bibinfo {author} {\bibfnamefont {R.}~\bibnamefont {Liu}}, \bibinfo {author}
  {\bibfnamefont {B.}~\bibnamefont {Divinskiy}}, \bibinfo {author}
  {\bibfnamefont {A.}~\bibnamefont {Telegin}}, \ and\ \bibinfo {author}
  {\bibfnamefont {S.~O.}\ \bibnamefont {Demokritov}},\ }\href {\doibase
  10.1038/ncomms10446} {\bibfield  {journal} {\bibinfo  {journal} {Nature
  Communications}\ }\textbf {\bibinfo {volume} {7}} (\bibinfo {year} {2016}),\
  10.1038/ncomms10446}\BibitemShut {NoStop}%
\end{thebibliography}%

\end{document}